\pdfoutput=1
\documentclass[11pt]{article}
\usepackage{amssymb}
\usepackage{graphicx}
\usepackage{pifont}
\parskip \baselineskip
\newcommand{\s}{\subseteq}

\newcommand{\sm}{\bf \stackrel{.\ .}{\smile}}
\newcommand{\HH}{\mathbb{H}}

\newcommand{\ol}{\overline}

\newcommand{\ra}{\rightarrow}
\newcommand{\es}{\emptyset}
\newcommand{\Ra}{\Rightarrow}
\newcommand{\LRa}{\Leftrightarrow}
\newcommand{\HAH}{{\it Horn$\wedge$AntiHorn} }

\begin{document}

\title{Polynomial-time satisfiability for a special case  of {\tt  Positive$\wedge$Negative}}

\author{Marcel
Wild\\[3pt]
Department of Mathematical Sciences,
University of Stellenbosch, South Africa\\
mwild@sun.ac.za, ORCID: 0000-0002-9773-1920
}

\date{}
\maketitle

\begin{quote}
A{\scriptsize BSTRACT}: {\footnotesize  A Boolean function in CNF format is of type {\tt  Positive$\wedge$Negative} if each  clause $C$ is either positive (i.e. all literals of $C$ are  positive) or negative (i.e. all  literals of $C$ are  negative). As is well known, deciding the satisfiability of such CNFs is NP-complete. We say that a CNF is of type {\tt DisjointPositive} 
if  its clauses are positive and  mutually disjoint. Dually define {\tt DisjointNegative}. It is shown that the satisfiability of CNFs of type {\tt  DisjointPositive$\ \wedge\ $DisjointNegative} can be decided in quadratic time.
Moreover, the modelset can be output in polynomial total time. This is relevant since it affects not only the modelsets of CNFs of type {\tt  Positive$\wedge$Negative}, but more generally of type {\tt  Horn$\wedge$AntiHorn}. As to the latter CNFs, they e.g. occur in connection with the fixpoints of a  Monotone Boolean Network. In another vein, the unsatisfiability of a Horn$\wedge$AntiHorn CNF can be demonstrated by means wholly different to the often used method of clausal proofs.}

\vspace{5mm}

\end{quote}

\section{Introduction}

Let us start by rephrasing and complementing an observation of Richard Goodman about the present article: There is an old asymmetry at the heart of automated reasoning. When a satisfiability solver answers {\it yes}, it can hand you the satisfying assignment itself. When the solver answers {\it no}, how can it certify\footnote {For instance the unsatisfiability-certificate in [HK] is 200 terabytes heavy and took 4 CPU years. More about [HK] in Subsection 8.5.} this? For a fragment of type {\tt Positive$\wedge$Negative} CNF's it is the other way around. If the answer is {\it no}, this can always be proven at once by pinpointing a 0-1-clash. Yet if the answer is {\it yes} then producing a satisfying assignment is comparatively hard. (Here "hard" mostly concerns the justification of steps, rather than the quadratic time cost.) What is more, {\it every} CNF splits with modest effort in fragments of the mentioned type (although, admittedly, there may be many fragments).

We assume a basic familiarity with Boolean functions. Thus recall that {\it clauses} $C$ are Boolean formulas of particular simplicity, i.e. by definition $C$ is a disjunction of literals such as $x_1\vee x_2\vee x_3\vee\ol{x}_4$. 
The {\it length} of a clause is the number of literals appearing in it. Recall that
 a {\it Conjunctive Normal Form (CNF)} is a conjunction of clauses such as
$$(1)\quad (x_1\vee x_2\vee x_3\vee\ol{x}_4)\wedge (\ol{x}_2\vee x_4\vee\ol{x}_5\vee x_6)\ :=\ C_1\wedge C_2$$
Let $f:\{0,1\}^m\to\{0,1\}$ be any\footnote{There are various ways to define Boolean functions. For us $f$ will mostly be defined by a CNF.} Boolean function. Then a {\it bitstring}, i.e. member $y\in\{0,1\}^m$, is a {\it model} of $f$ if $f(y)=1$. We write $Mod(f)$ for the set of all models. As is well known, if $f$ is rendered by a CNF, then it is NP-complete to decide whether $f$ is {\it satisfiable} (i.e. whether $Mod(f)\neq\es$).

Recall from the Abstract that a clause is {\it positive (negative)} if all its literals are  
positive (negative).
Further, a clause $C$ is a {\tt Horn-clause} (briefly: $C$ is {\tt Horn}), if it has at most one positive literal.
 Likewise $C$ is {\tt AntiHorn} if it has at most one negative literal (such as $C_1$ in (1)). Observe that $C_2$ in (1) is neither {\tt Horn} nor {\tt AntiHorn}. Going one level up, a CNF is a {\tt Horn-CNF} if all its clauses are {\tt Horn}. Likewise a CNF is a {\tt AntiHorn-CNF} if all its clauses are {\tt AntiHorn}.
As is well known, the satisfiability of either type  can be decided in linear time.
 A {\tt Horn-CNF} with {\it only} negative clauses is called a {\tt Negative-CNF}. 
 Dually, an {\tt AntiHorn-CNF} with {\it only} positive clauses is called a {\tt Positive-CNF}. 
 
  The CNF $f$ is said to be (of type) {\tt Horn$\wedge$AntiHorn} if it can be written as $f=f_1\wedge f_2$ such that $f_1$ is a {\tt Horn-CNF} and $f_2$ is a {\tt AntiHorn-CNF}. As opposed to {\tt Horn} and {\tt AntiHorn} {\it individually}, deciding the satisfiability of   {\tt Horn$\wedge$AntiHorn} CNFs is NP-complete.
In fact, already {\tt Positive$\wedge$Negative} is NP-complete. Nevertheless, our core result (Theorem 1) states that deciding satisfiability in quadratic time is possible for some relevant special case of {\tt Positive$\wedge$Negative}.

{\bf 1.1} Here comes the Section break-up. Section 2 will  survey itself  at  the beginning. In fact, since Section 2 may distract from the overall storyline, the reader is advised to skip Section 2 at a first reading and come back to it when its content is quoted in later Sections.

As to Section 3,
by definition a {\tt DisjointPositive-CNF} is a {\tt Positive-CNF} whose clauses are mutually disjoint, i.e. distinct clauses have no common literal. We show that the modelsets of {\tt DisjointPositive-CNFs}
 are neatly captured by so-called {\it 012e-rows} $\rho$. (Here  "2" is the familiar don't-care symbol but the "e" symbol, due to the author, is less known.)
  In a dual fashion one defines {\tt DisjointNegative-CNFs} and {\it 012n-rows} $\sigma$.
  We state in Theorem 1 that the satisfiability of type  {\tt DisjointPositive$\wedge$DisjointNegative} CNF's, i.e. the non-emptyness of $\rho\cap\sigma$, can be decided in linear time (the proof being postponed to Section 9).  While the principle of inclusion-exclusion (PIE)  decides $\rho\cap\sigma\stackrel{?}{=}\es$ in more familiar ways (and gives the precise cardinality $|\rho\cap\sigma|$ as a perk), unfortunately the PIE takes exponential time.
Furthermore, the concrete members (=bitstrings) in $\rho\cap\sigma$ remain utterly unknown. 

Sections 4,5,6,7 all revolve around $\rho\cap\sigma\stackrel{?}{=}\es $ and hopefully increase the reader's acceptance (or even appetite) for the technicalities of Section 9. Section 4 records well known facts about closure systems, set-ideals, set-filters, and various ways to capture them. Section 5 touches upon the notorious problem to enumerate the minimal hitting sets of a hypergraph  in polynomial total time. Section 6 is about the classic Ramsey numbers $R(k,\ell)$ and shows that $n<R(k,\ell)$ is equivalent to the existence of a suitable 012e-row $\rho$, and a suitable 012n-row $\sigma$ such that
$\rho\cap\sigma\neq\es$. While testing $\rho\cap\sigma\stackrel{?}{=}\es $ is fast due to Theorem 1, the problem that remains in Section 6 is to reduce  the large pools of $\rho_i$'s and $\sigma_j$'s in which $\rho$ and $\sigma$ are to be found.
 Hints are provided but much work remains to be done.
Section 7 features  a proper {\tt Horn$\wedge$AntiHorn} scenario (not just {\tt Positive$\wedge$Negative} as in Sections 5,6)
that involves Monotone Boolean Networks, i.e. certain functions $\Phi:\{0,1\}^m\to\{0,1\}^m$. Namely, we show that the fixpoints $y$ of $\Phi$ (i.e. $\Phi(y)=y$) are exactly the models of a readly calculated CNF $g$ of type {\tt Horn$\wedge$AntiHorn}. Hence  the $\rho,\sigma$-mechanism is applicable to enumerate all fixpoints.

 Section 8 sticks to {\tt Horn$\wedge$AntiHorn} and shows that {\it each} CNF $f$ (not necessarily related to Boolean networks) is equisatisfiable\footnote{Generally two Boolean functions $f,g$ are {\it equisatisfiable} when $f$ is satisfiable iff $g$ is satisfiable. (This implies nothing about their actual satisfiability)} with a certain CNF $g$ of type {\tt Horn$\wedge$AntiHorn}.  The modelset $Mod(g)\s\{0,1\}^{m+s}$ (which hence can be calculated with the $\rho,\sigma$-mechanism) yields the modelset of interest, i.e. $Mod(f)\s\{0,1\}^{m}$, as the {\it projection} of $Mod(g)$ onto the first $m$ variables. (The difference between $m$ and $m+s$ is small.) Put another way, the  fragments (alluded to at the beginning of Sec.1) into which each CNF can be partitioned, get unravelled in Section 8 as set-systems of type $\rho\cap\sigma$.

 Interestingly, 3-CNFs show up in both Section 7 and 8 at various places.

As to Section 9, let $\rho$  be any $012e$-row and $\sigma$ any  $012n$-row of the same length $m$. We provide the pending proof of Theorem 1, i.e. we show that $\rho\cap\sigma\stackrel{?}{=}\es$
can be decided in time $O(m^2)$.  This is the crucial ingredient for Theorem 5 which establishes that in fact {\it all} bitstrings in $\rho\cap\sigma$ can be listed in output\footnote{We sometimes use the synonym "polynomial total time".} polynomial time.

Without further mention, all sets in this article  are assumed to be {\bf finite}.

\section{Making 012-rows disjoint}

As previously mentioned, Section 2 can be skipped at a first reading; more precisely the skipping concerns the more technical\footnote{Although quite technical, Section 2 only deals with plain 0-bits, 1-bits, and don't-care symbols. This contrasts with later Sections that  use "baroque" (yet highly efficient) symbols to capture sets of bitstrings. Apart from catering for Sections  4 and 7, Section 2 is actually of broader interest.} Subsections 2.1 and 2.2.

First off, we will often identify {\it bitstrings} $y\in\{0,1\}^m$ with their supports $\{i: y_i=1\}$, i.e. with subsets of $\{1,2,..,m\}$.
By definition a {\it 012-row} such as $r:=(0,2,1,0,2)$ represents the set of bitstrings (subsets) 
$$\Big\{(0,{\bf 0},1,0,{\bf 0}),(0,{\bf 0},1,0,{\bf 1}),(0,{\bf 1},1,0,{\bf 0}),(0,{\bf 1},1,0,{\bf 1})\Big\}=\Big\{\{3\},\{3,5\},\{2,3\},\{2,3,5\}\Big\}.$$ Thus "2" is the familiar\footnote{Many texts use "$\ast$" instead of "2".} don't-care symbol that can freely be replaced by a 0-bit or a 1-bit. We put $zeros(r):=\{1,4\},\ ones(r):=\{3\},\ twos (r):=\{2,5\}$, and likewise for arbitrary 012-rows $r'$. From the above it is evident that $|r'|=2^\alpha$, where $\alpha:=|twos(r')|$.
It is also clear that the {\it intersection} of same length 012-rows is either empty
(due to 0-1 {\it clashes} as in $(0,1,2)\cap(2,0,2)=\es$), or can otherwise again be written as 012-row:
$(2,0,1,1,2,2)\cap(0,2,2,1,2,0)=(0,0,1,1,2,0)$.

 As to {\it unions} of 012-rows, how large is
$$r_1\cup r_2\cup r_3:=(2,2,2,2,0,1,0)\cup(0,0,2,2,2,1,2)\cup(2,2,0,0,0,2,0)\ ?$$
Inclusion-exclusion yields the cardinality
$$\quad|r_1\cup r_2\cup r_3|=|r_1|+|r_2|+|r_3|-|r_1\cap r_2|-|r_1\cap r_3|-|r_2\cap r_3|+|r_1\cap r_2\cap r_3|$$
$$=\ 16+16+8-4-4-1+1\ =\ 32$$
right away, but of course the workload rises exponentially with the number of 012-rows involved. 

{\bf 2.1} If one could readily replace a union of $t$ many 012-rows by a {\it disjoint} union of "slightly" more (say $p$ many) 012-rows, this would beat inclusion-exclusion since then $p$ instead of $2^t-1$ numbers need to be added. Furthermore in many scenarios  not just their number, but  the  bitstrings {\it themselves} are important. Listing them without\footnote{We henceforth use $\uplus$ to indicate disjoint union.} overlaps (i.e. $\rho'_1\uplus\cdots\uplus\rho'_p$)
is certainly more helpful than the original offering (i.e $r_1\cup\cdots\cup r_t$).

\begin{tabular}{l|c|c|c|c|c|c|c|l}
\hline
	${r_1}:=$ &  2 & 2 & 2 & 2 & 0& 1 & 0 \\ \hline
    ${r_2}:=$ & 0 & 0 & 2 & 2 & 2& 1 & 2 \\ \hline
    & & & & & & &   \\ \hline
    ${\rho_1}:=$ &  \bf 1 & \bf 2 & 2 & 2 & 0& 1 & 0 \\ \hline
    ${\rho_2}:=$ & \bf 0 & \bf 1 & 2 & 2 & 0& 1 & 0 \\ \hline
     & & & & & & &   \\ \hline
    ${r_3}:=$ &  2&  2 & 0 & 0 & 0& 2 & 0 \\ \hline
     & & & & & & &   \\ \hline
      ${\rho_3}:=$ &   1 &  2 & \bf 1 & \bf 2 & 0& 1 & 0 \\ \hline
      ${\rho_4}:=$ &   1 &  2 & \bf 0 & \bf 1 & 0& 1 & 0 \\ \hline
    ${\rho_5}:=$ &  0 &  1 & \bf 1 &\bf 2 & 0& 1 & 0 \\ \hline
    ${\rho_6}:=$ & 0 &  1 & \bf 0 &\bf 1 & 0& 1 & 0 \\ \hline
      & & & & & & &   \\ \hline
      ${\rho_7}:=$ & 0 & 0 & \bf 1 &\bf 2 &\bf 2& 1 &\bf 2 \\ \hline
       ${\rho_8}:=$ & 0 & 0 & \bf 0 &\bf 1 &\bf 2& 1 &\bf 2 \\ \hline
        ${\rho_9}:=$ & 0 & 0 & \bf 0 &\bf 0 &\bf 1& 1 &\bf 2 \\ \hline
         $\rho_{10}:=$ & 0 & 0 & \bf 0 &\bf 0 &\bf 0& 1 &\bf 1 \\ \hline
\end{tabular}

{\sl Table 1: Making the union $r_1\cup r_2\cup r_3$ disjoint}

 Evidently $r_1\cup r_2 =(r_1\setminus r_2)\uplus r_2$. 
It remains to represent $Y:=r_1\setminus r_2$ as disjoint union of 012-rows. If (as here) neither the trivial case $Y=r_1$ nor $Y=\es$ occurs, then (i) 0-1 clashes are absent and (ii) there {\it are} bitstrings $y\in Y$. One option for $y\in r_1$ to "detach" itself from the "crowd" in $r_2$ is to have 1-bits at places where all bitstrings in $r_2$ must have 0-bits (i.e. $y_i=1$ for some $i\in zeros(r_2)$). Clearly $\rho_1\uplus\rho_2$ is the set of $y\in r_1$ that exploit this option. A moment's thought confirms that, 0-1 clashes being absent, mentioned option is the {\it only}\footnote{In the more general scenario of 4.6.3 there will be a second option.} option to detach yourself from the crowd.
In other words, $r_1\setminus r_2=\rho_1\uplus\rho_2$, and so $r_1\cup r_2=\rho_1\uplus\rho_2\uplus r_2$. Therefore
$$[r_1\cup r_2]\cup r_3=[(\rho_1\setminus r_3)\uplus (\rho_2\setminus r_3)\uplus (r_2\setminus r_3)]\uplus r_3.$$
Using detachment to expand each of $\rho_1\!\setminus\! r_3,\ \rho_2\!\setminus\! r_3,\ r_2\!\setminus\! r_3$ into  012-rows,  yields (see Table 1 and Subsection 2.2)
$$r_1\cup r_2\cup r_3=(\rho_3\uplus\rho_4)\uplus(\rho_5\uplus\rho_6)\uplus (\rho_7\uplus\rho_8\uplus \rho_9\uplus\rho_{10})\uplus r_3.$$
One checks that the cardinalities of $\rho_3,...,\rho_{10},r_3$ sum up to 32, in accordance with inclusion-exclusion.

\begin{center}
   \includegraphics[scale=0.8]{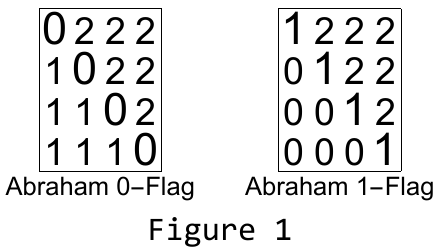} 
\end{center}

{\bf 2.2} A few comments are in order concerning the more subtle expansion of $r_2\setminus r_3$. In its core  detachment boils down to find the set $Y$ of all  $y\in (2,2,...,2)$ ($n$ components) that detach themselves from $(0,0,...,0)$. Of course $Y=(1,0,...,0)\cup (0,1,...,0)\cup\cdots\cup(0,0,...,1)$. This union can be made disjoint by using an Abraham\footnote{They were introduced, in honor of J.A. Abraham, in [W1,p.1078].} 1-Flag of dimension $n\times n$. Figure 1 pictures a $4\times 4$ Abraham 1-Flag. One checks that its four 012-rows are disjoint and their union is $(2,2,2,2)\setminus\{(0,0,0,0)\}$. The cardinalities behave accordingly: $8+4+2+1=2^4-1$. It is clear how the $4\times 4$ pattern extends to the
$n\times n$ pattern, and accordingly $2^{n-1}+2^{n-2}+\cdots+2+1=2^n-1$. Abraham 1-Flags, and dually Abraham 0-Flags (also Figure 1), will accompany us throughout this article.

\section{Definition and basic properties of 012e-rows and 012n-rows}

Recall from the Introduction that {\tt DisjointPositive-CNFs} are {\tt Positive-CNFs} with disjoint clauses.  Let us add any number of {\bf length 1 negative} clauses. The result we\\ call\footnote{The possible impression that this little extra generality is unwarranted will soon be corrected.
Of course, in general the indices of the negative literals (here 1,2) are arbitrary.}
{\tt DisjointPossitive}, an example being $f_0$ in (2). 
Put another way, dropping the boldface entries in (2) brings us from {\tt DisjointPossitive} to {\tt DisjointPositive}:

$$(2)\quad f_0={\bf \ol{x}_1\wedge\ol{x}_2}\wedge x_3\wedge x_4\wedge (x_6\vee x_7\vee x_8)\wedge (x_9\vee x_{12})\wedge(x_{10}\vee x_{11})\wedge (x_{13}\vee x_{14}\vee x_{15})$$

{\bf 3.1}
With the {\tt DisjointPossitive-CNF} in (2) we associate the {\it 012e-row} $\rho_0$  shown in $(2')$. Specifically, the  negative length 1 clauses $\ol{x}_1,\ol{x}_2$ are represented by the 0's in positions 1 and 2 of $\rho_0$. The
 positive length 1 clauses $x_3,x_4$ are represented by the 1's in positions 3 and 4 of $\rho_0$. 
$$(2')\quad \rho_0:=(0,0,1,1,2,e_1,e_1,e_1,e_3,e_2,e_2,e_3,e_4,e_4,e_4)$$
Each clause of length $\ge 2$ gets its own {\it e-wildcard} $(e_i,e_i,..,e_i)$, distinct clauses being distinguished by subscripts. For instance  $x_9\vee x_{12}$ is represented by the entries $e_3,e_3$ located at the positions 9 and 12. 
The don't-care "2" at position 5 signifies that neither $x_5$ nor $\ol{x}_5$ occur in $f_0$.
The following acronyms are now self-explanatory:
$$ones(\rho_0)=\{3,4\},\ zeros(\rho_0)=\{1,2\},\ twos(\rho_0)=\{5\},\ pos(e_3,e_3)=\{9,12\}$$

By definition, an arbitrary 012e-row of length $m$  (e.g. $m=15$ for $\rho_0$) represents the following set $S(\rho)\s\{0,1\}^m$ of bitstrings. It holds that $y\in\{0,1\}^{m}$ belongs to $S(\rho)$ iff these conditions hold:
\begin{itemize}
    \item[(i)] $zeros(\rho)\s zeros(y)$
     \item[(ii)] $ones(\rho)\s ones(y)$
     \item[(iii)] {\it $ones(y)\cap pos(e_i,...,e_i)\neq\es$ or all e-wildcards $(e_i,..,e_i)$ of $\rho$.}
\end{itemize}
It follows that 012e-rows are ideally\footnote{Notice that a DisjointPossitive-CNF like $(x_1\vee x_2\vee x_3)\wedge (x_4\vee x_5)\wedge \ol{x}_2\wedge\ol{x}_6$ is equivalent to the smoother DisjointPossitive-CNF $(x_1\vee x_3)\wedge (x_4\vee x_5)\wedge\ol{x}_6$. It is hence no extra restriction (but is convenient) if we henceforth assume that this kind of smoothing has been carried out already.} suited to represent the modelsets of {\tt DisjointPossitive-CNFs}, e.g. $Mod(f_0)=S(\rho_0)$. As to "represent", we will henceforth be more sloppy and simply write $Mod(f_0)=\rho_0$. Strictly speaking, this confounds a "string of symbols" with a "set of bitstrings", yet from the context it will always be clear which of the two is meant.

    {\bf 3.2} Dually one defines {\tt DisjointNeggative-CNFs}, an example being $g_0$ in (3).
$$(3)\quad g_0=(\ol{x}_2\vee\ol{x}_7)\wedge(\ol{x}_{3}\vee \ol{x}_{13}\vee\ol{x}_{14})\wedge(\ol{x}_4\vee\ol{x}_5\vee\ol{x}_9\vee\ol{x}_{11}\vee\ol{x}_{15})\wedge(\ol{x}_6\vee\ol{x}_8)\wedge  \ol{x}_{10}\wedge {\bf x_{12}}\ $$

Dropping the boldface entry in (3) brings us from {\tt DisjointNeggative} to {\tt DisjointNegative}.
      In dual fashion the {\tt DisjointNeggative-CNF} $g_0$ in (3) triggers this {\it 012n-row} which uses {\it n-wildcards} such as $(n_4,n_4,n_4)$:
$$(3')\quad \sigma_0:=(2,n_1,n_4,n_3,n_3,n_2,n_1,n_2,n_3,0,n_3,1,n_4,n_4,n_3)$$
It e.g. holds that $pos(n_3,..,n_3)=\{4,5,9,11,15\}$. Dual to the above, a 012n-row $\sigma$ of length $m$ represents the following set $S'(\rho)\s\{0,1\}^m$ of bitstrings. It holds that $y\in\{0,1\}^{m}$ belongs to $S'(\sigma)$ iff these conditions hold (notice how (iii') differs from (iii)):
\begin{itemize}
    \item[(i')] $zeros(\sigma)\s zeros(y)$
     \item[(ii')] $ones(\sigma)\s ones(y)$
     \item[(iii')] {\it $zeros(y)\cap pos(n_i,...,n_i)\neq\es$ or all n-wildcards $(n_i,..,n_i)$ of $\sigma$.}
\end{itemize}
Analogous to 3.1  we often identify $S'(\sigma)$ and $\sigma$; thus it e.g. holds (check) that $Mod(g_0)=\sigma_0$.

{\bf 3.3} The most obvious reason for a 012e-row $\rho$ to have no bitstring in common with a same length 012n-row $\sigma$ are  0-1 clashes, which we met already in Section 2. Formally this happens iff either $ones(\rho)\cap zeros(\sigma)=\es$ or $ones(\sigma)\cap zeros(\rho)=\es$.
 Apart from 0-1 clashes either of the following {\it trivial reasons} is sufficient for $\rho\cap\sigma=\emptyset$:
\begin{itemize}
	\item There is an e-wildcard $(e,..,e)$ of $\rho$ with $pos(e,..,e)\s zeros(\sigma)$
	\item There is a n-wildcard $(n,..,n)$ of $\sigma$ with $pos(n,..,n)\s ones(\rho)$
\end{itemize}

Yet the emptyness of $\rho\cap\sigma$ can have {\it non-trivial} reasons. For instance

$\rho\cap\sigma\ :=\ (1,1,e,e)\cap (n,n',n,n')=\Big( (1,1,e,e)\cap (1,1,2,2)\Big)\cap (n,n',n,n')\\  \hspace*{1.4cm}= (1,1,e,e)\cap \Big( (1,1,2,2)\cap (n,n',n,n')\Big)=(1,1,e,e)\cap (1,1,0,0)=\emptyset.$

{\bf 3.4} For general $\rho\cap\sigma$ the trick above (associativity of set intersection) won't get us far. Instead we will reduce $\rho:=\rho_0$ to $\rho_0\supseteq \rho_1
\supseteq\cdots\supseteq\rho_t$ and $\sigma:=\sigma_0$ to $\sigma_0\supseteq \sigma_1
\supseteq\cdots\supseteq\sigma_t$ such that 
 $\rho_i\cap\sigma_i=\rho_0\cap\sigma_0$ for all $0\le i\le t$. Exactly one of two things will occur. {\bf Case A:} $\rho_i\cap\sigma_i=\es$ for some index $i$, due to trivial reasons. Then we stop, knowing that $\rho_0\cap\sigma_0=\es$.
  {\bf Case B:} No trivial reasons occur. Then our reduction process stabilizes with $\rho_t$ and $\sigma_t$.

 To explain what exactly is meant by "reduce" and "stabilize", consider $\rho_0$ and $\sigma_0$ in Table 2 (they coincide with the rows in $(2')$ and $(3')$):

\begin{tabular}{l|c|c|c|c|c|c|c|c|c|c|c|c|c|c|c|c}
	
	&1 &2 &3 &4 &5 &6 &7 &8 &9 &10 &11 &12 &13 &14 &15   \\ \hline
	& & & & & & & & & & & & & & & &\\ \hline	
	$\rho_0=$ &  0 &  0 &  1 &  1 &  2 &  $e_1$ &$e_1$  &  $e_1$&  $e_3$ &  $e_2$ &  $e_2$ &  $e_3$ &  $e_4$ &$e_4$ &$e_4$   \\ \hline
	$\sigma_0=$ &  2 &  $n_1$ &  $n_4$ &  $n_3$ &  $n_3$ &  $n_2$ &$n_1$  &  $n_2$&  $n_3$ &  0 &  $n_3$ &  1 &  $n_4$ &$n_4$ &$n_3$   \\ \hline
	& & & & & & & & & & & & & & & &\\ \hline	
	$\rho_1=$ &  0 &  0 &  1 &  1 &  2 &  $e_1$ &$e_1$  &  $e_1$&  $e_3$ &  $e_2$ &  $e_2$ &  $e_3$ &  $e_4$ &$e_4$ &$e_4$   \\ \hline
	$\sigma_1=$ &  {\bf 0} &   {\bf 0} &   {\bf 1} &   {\bf 1} &  $n_3$ &  $n_2$ &$\ra 2$  &  $n_2$&  $n_3$ &  0 &  $n_3$ &  1 &  $n_4$ &$n_4$ &$n_3$   \\ \hline
	& & & & & & & & & & & & & & & &\\ \hline	
	$\rho_2=$ &  0 &  0 &  1 &  1 &  2 &  $e_1$ &$e_1$  &  $e_1$&  $\ra 2$ &  {\bf 0} &$\ra 1$ & {\bf 1} &  $e_4$ &$e_4$ &$e_4$   \\ \hline
	$\sigma_2=$ &   0 &    0 &    1 &   1 &  $n_3$ &  $n_2$ &2  &  $n_2$&  $n_3$ &  0 &  $n_3$ &  1 &  $n_4$ &$n_4$ &$n_3$   \\ \hline
	& & & & & & & & & & & & & & & &\\ \hline	
	$\rho_3=$ &  0 &  0 &  1 &  1 &  2 &  $e_1$ &$e_1$  &  $e_1$&  2 &   0 &  1 &  1 &  $e_4$ &$e_4$ &$e_4$   \\ \hline
	$\sigma_3=$ &   0 &    0 &    1 &   1 &  $n_3$ &  $n_2$ &2  &  $n_2$&  $n_3$ &  0 &  {\bf 1} &  1 &  $n_4$ &$n_4$ &$n_3$   \\ \hline	
\end{tabular}

{\sl Table 2: Reducing type $(012e,012n)$ intersections to type $(2e,2n)$ intersections}

We obtain $\sigma_1$ by enforcing\footnote{Generally newly enforced  components $\bf 0,1$ are rendered boldface. They trigger further adaptions:  $(n_1,n_1)$ in $\sigma_0$ becomes $(0,2)$ in $\sigma_1$, and $(n_3,n_3,n_3,n_3)$ becomes $(1,n_3,n_3,n_3)$,
	and $(n_4,n_4,n_4)$ becomes $(1,n_4,n_4)$.  The repercussions of forced 0's and 1's are rendered as $\ra 2$ or $\ra 1$ or $\ra 0$. The latter will occur in 3.4.1.}	  the 0's and 1's of $\rho_1:=\rho_0$ upon $\sigma_0$. Although $\sigma_1\subset \sigma_0$ (proper inclusion), the reader should convince himself that $\rho_1\cap\sigma_1=\rho_0\cap\sigma_0$. Likewise we obtain $\rho_2$ by enforcing the 0's and 1's of $\sigma_1\ (=:\sigma_2)$ upon $\rho_1$. Again $\rho_2\subset\rho_1$ yet
$\rho_2\cap\sigma_2=\rho_1\cap\sigma_1$. After one more step we get $\rho_3\subseteq\rho_2$   and
$\sigma_3\subseteq\sigma_2$ (in fact $\rho_3:=\rho_2$ and $\sigma_3\subset\sigma_2$) such that $\rho_3\cap\sigma_3=\rho_2\cap\sigma_2=\cdots=\rho_0\cap\sigma_0$.
The reduction process stops because now $ones(\rho_3)=ones(\sigma_3)$ and $zeros(\rho_3)=zeros(\sigma_3)$.
Putting $\ol{\rho_3}=(2,e_1,e_1,e_1,2,e_4,e_4,e_4)$
and $\ol{\sigma_3}=(n_3,n_2,2,n_2,n_3,n_4,n_4,n_3)$ (both indexed by $5,6,7,8,9,13,14,15$), it is clear that $\rho_0\cap\sigma_0=\es$ iff $\ol{\rho}_3\cap\ol{\sigma}_3=\es$. This is progress because we managed to get rid of the 0-bits and 1-bits. Specifically, if we find a bitstring $y'\in\ol{\rho}_3\cap\ol{\sigma}_3$ and inject  three 0's and four 1's at the right places, we get a bitstring $y\in\rho_0\cap\sigma_0$.

 Generally let  $\rho_0$ and $\sigma_0$ be length $m$ rows of type 012e and 012n respectively. As illustrated in Table 2, triggered by $(\rho_0,\sigma_0)$ one calculates $(\rho_1,\sigma_1),(\rho_2,\sigma_2)$, and so forth until one either finds that
$\rho_0\cap\sigma_0=\es$ due to trival reasons, or else the sketched 0,1,2-{\it propagation} stops at $(\rho_t,\sigma_t)$ (thus $t=3$ in Table 2).

{\bf 3.4.1} Before we embark on the time assessment, here comes an example of the other kind (Table 2'). Starting with the pair $(\rho_0',\sigma_0')$, the $0$ of $\sigma'_0$ inflicts itself on $\rho'_1$ (=successor of $\rho'_0$), thereby triggering a 1-bit as well (because of $e_1,e_1$ in $\rho'_0$). In a similar fashion $\rho'_1$ triggers $\sigma'_1$. By definition of $\rho'_1,\sigma'_1$, every potential bitstring in $\rho_0'\cap\sigma_0'$ belongs to both
$\rho_1'$ and $\sigma_1'$. In fact it will belong to both
$\rho_i'$ and $\sigma_i'\ (i\ge 1)$ as we go on this way. In our example, the following happens for $i=3$: Both $\rho'_3$ and $\sigma'_3$ boil down to 01-rows (=bitstrings), but to different ones! This shows that $\rho_0'\cap \sigma_0'=\es$.

\begin{tabular}{l|c|c|c|c|c|c|c|l}
\hline
	${\rho'_0}:=$ &  $e_1$ & $e_1$ & $e_2$ & $e_2$ & $e_3$& $e_3$ & 1 \\ \hline
    ${\sigma'_0}:=$ &  $0$ & $n_1$ & $n_1$ & $n_2$ & $n_2$& $n_3$ & $n_3$ \\ \hline
    ${\rho'_1}:=$ &  $\bf 0$ & $\ra 1$ & $e_2$ & $e_2$ & $e_3$& $e_3$ & 1 \\ \hline
    ${\sigma'_1}:=$ &  $0$ & $\bf 1$ & $\ra 0$ & $n_2$ & $n_2$& $n_3$ & $n_3$ \\ \hline
     ${\rho'_2}:=$ &  $0$ & $1$ & $\bf 0$ & $\ra 1$ & $e_3$& $e_3$ & 1 \\ \hline
      ${\sigma'_2}:=$ &  $0$ & $1$ & $0$ & $\bf 1$ & $\ra 0$& $n_3$ & $n_3$ \\ \hline
      ${\rho'_3}:=$ &  $0$ & $1$ & $ 0$ & $1$ & $\bf 0$& $\ra 1$ & 1 \\ \hline
       ${\sigma'_3}:=$ &  $0$ & $1$ & $0$ & $1$ & $0$& $\bf 1$ & $\ra 0!$ \\ \hline
\end{tabular}

{\sl Table 2': Proving type $(012e,012n)$ intersections to be empty}

{\bf 3.5} In order to access the cost $O(..)$ of these manipulations we must specify more carefully the data structure underlying the starter rows $\rho_0,\sigma_0$ (and their successors). Looking at $\rho_0$ (dually for $\sigma_0$), it suffices to update\footnote{For instance, $ones$ starts off as $ones(\rho_0)$, then becomes $ones(\rho_1)$, and so forth. Note that a parameter $twos$ is superfluous;  at any moment it is implicitely given as the complement of $ones\uplus zeros\uplus pos(e_1)\uplus\cdots$. For $\sigma_0$ everything works dually, using $ones',zeros',pos(n_1), etc.$  } the parameters $ones,zeros,pos(e_1),pos(e_2)$, etc. 

Suppose that 0,1,2-propagation puts 1 on some $e_2$ symbol at position $j$. This merely triggers $ones:=ones\cup\{j\}$ (programmer's talk) and $pos(e_2):=\es$, and so has constant complexity $O(1)$. Slightly more cumbersome, suppose that 0,1,2-propagation puts 0 on some $e_7$ symbol at position $k$.
This  forces $zeros:=zeros\cup\{k\}$ and leads to  two subcases. If $|pos(e_7)|\ge 3$, then $pos(e_7):=pos(e_7)\setminus\{k\}$. If $pos(e_7)=\{k,\ell\}$ has only two elements, then $ones:=ones\cup\{\ell\}$ and $pos(e_7):=\es$. The first subcase is more expensive and costs\footnote{Using  clever data structures (tree-based or hash-based sets) one could push $O(|pos(e_7)|)$ to $O(log|pos(e_7)|)$ or even further. But this is pointless in view of some $O(m^2)$ cost around the corner.} $O(|pos(e_7)|)=O(m)$. At most $m$ many e-components get changed on the journey from $\rho_0$ to $\rho_t$. Likewise for $\sigma_0$. Hence the cost incurred so far is $O(m^2)$.

However this is not all. Testing repeatedly for 0-1 clashes, i.e. testing $ones\cap zeros'\stackrel{?}{=}\es$
and $ones'\cap zeros\stackrel{?}{=}\es$, costs $O(m)$. Since "repeatedly" can only be bound by "$\le m$", this accumulates to $O(m^2)$. Fortunately $O(m^2)$ also covers the other  trivial reasons (see 3.3) for $\rho_i\cap\sigma_i=\es$, as well as the above  "cost incurred so far".  It follows that the overall cost of our reduction to 2e- and 2n-level is $O(m^2)$.

{\bf 3.6} We see that deciding the satisfaction of {\tt DisjointPositive$\hspace{2pt} \wedge\hspace{2pt}$DisjointNegative} boils down to deciding the emptiness of $\rho\cap\sigma$, where $\rho$ is a 2e-row and $\sigma$ a 2n-row of  the same length. Surprisingly, it turns out that {\it always} $\rho\cap\sigma\neq\es$, and so:

{\bf Theorem 1:} {\it Let $f:\{0,1\}^m\to\{0,1\}$ be a Boolean function which is in CNF format and of type
{\tt DisjointPossitive$\hspace{2pt} \wedge\hspace{2pt} $DisjointNeggative}. Then the satisfiability of $f$ can be tested in $O(m^2)$ time.}

The argument that always $\rho\cap\sigma\neq\es$ is quite\footnote{For the impatient reader: the gist of it can be gleaned from 9.4.1.} subtle and is postponed to Section 9. In the remainder of Section 3 we continue with  elementary observations about $2e$-rows and $2n$-rows.

{\bf 3.7}
Let $\rho$ and $\sigma$ be 2e- and 2n-rows respectively (always of the same length).
We just mentioned that always $|\rho\cap\sigma|>0$. But what is the precise value  of $|\rho\cap\sigma|$? The principle of inclusion-exclusion (PIE) will give the answer.
  Thus consider $\rho$ and $\sigma$ in Table 3.

\begin{tabular}{l|c|c|c|c|c|c|c|c|c|c|l}
	& 1 & 2 & 3& 4 & 5  & 6   & 7& 8& 9& cardinality & \\ \hline
	& & & & & & & & & & & \\ \hline
	${\sigma}:=$ &  $n_1$ & $n_1$ & $n_2$ & $n_2$ & $n_2$& $n_3$& $n_3$&  $n_3$& $2$ & $2\cdot 3\cdot 7^2=294$ \\ \hline
    ${\sigma(\stackrel{\rightarrow}{e_1})}:=$ &  2 & {\bf 0} & 2 & {\bf 0} & {\bf 0} & $n_3$& $n_3$&  $n_3$& $2$ & $2^3\cdot 7=56$ \\ \hline
     ${\sigma(\stackrel{\rightarrow}{e_2})}:=$ &  $n_1$ & $n_1$ & $n_2$ & $n_2$ & $n_2$ & 2& {\bf 0}&  {\bf 0}& {\bf 0} & $2\cdot 3\cdot 7=42$ \\ \hline
      ${\sigma(\stackrel{\rightarrow}{e_1},\stackrel{\rightarrow}{e_2})}:=$ &  2 & 0 & 2 & 0 & 0 & 2& 0&  0& 0 & $2^3=8$ \\ \hline
      & & & & & & & & & & & \\ \hline

      ${\rho}:=$ &  2 & $e_1$ & 2 & $e_1$ & $e_1$& 2 & $e_2$&  $e_2$& $e_2$ & $2^3\cdot 7^2=392$ \\ \hline
       ${\rho(\stackrel{\rightarrow}{n_1})}:=$ &  {\bf 1} & {\bf 1} & 2 & 2 & 2 & 2 & $e_2$&  $e_2$& $e_2$ & $2^4\cdot 7=112$ \\ \hline
        ${\rho(\stackrel{\rightarrow}{n_2})}:=$ &  2 & 2 & {\bf 1} & {\bf 1} & {\bf 1} & 2 & $e_2$&  $e_2$& $e_2$ & $2^3\cdot 7=56$ \\ \hline
        ${\rho(\stackrel{\rightarrow}{n_3})}:=$ &  2 & $e_1$ & 2 & $e_1$ & $e_1$ & {\bf 1} & {\bf 1}& {\bf 1}& 2 & $2^3\cdot 7=56$ \\ \hline
        ${\rho(\stackrel{\rightarrow}{n_1},\stackrel{\rightarrow}{n_2})}:=$ &  {\bf 1} & {\bf 1} & {\bf 1} & {\bf 1} & {\bf 1} & 2 & $e_2$& $e_2$& $e_2$ & $14$ \\ \hline
         ${\rho(\stackrel{\rightarrow}{n_1},\stackrel{\rightarrow}{n_3})}:=$ &  {\bf 1} & {\bf 1} & 2 & 2 & 2 & {\bf 1} & {\bf 1}& {\bf 1}& 2 & $16$ \\ \hline
         ${\rho(\stackrel{\rightarrow}{n_2},\stackrel{\rightarrow}{n_3})}:=$ &  2 & 2 & {\bf 1} & {\bf 1} & {\bf 1} & {\bf 1} & {\bf 1}& {\bf 1}& 2 & $8$ \\ \hline
         ${\rho(\stackrel{\rightarrow}{n_1},\stackrel{\rightarrow}{n_2},\stackrel{\rightarrow}{n_3})}:=$ &  {\bf 1} & {\bf 1} & {\bf 1} & {\bf 1} & {\bf 1} & {\bf 1} & {\bf 1}& {\bf 1}& 2 & $2$ \\ \hline             	
\end{tabular}

{\sl Table 3: Calculating $|\rho\cap\sigma|$ with the principle of inclusion-exclusion}

By definition $\sigma(\stackrel{\rightarrow}{e_1})$ is the set of all bitstrings $y\in\sigma$ that {\it violate} the wildcard $\stackrel{\rightarrow}{e_1}:=(e_1,e_1,e_1)$ of $\rho$ in the sense that $pos(\stackrel{\rightarrow}{e_1})\s zeros(y)$.
Likewise $\sigma(\stackrel{\rightarrow}{e_2})$
is defined, and $\sigma(\stackrel{\rightarrow}{e_1},\stackrel{\rightarrow}{e_2})$ is the set of $y$'s that violate both wildcards. Table 3 displays these three sets as 012n-rows.
Using the principle of inclusion-exclusion one concludes
$$ \big|\rho\cap\sigma\big|=\big|\sigma\big|-\big|\sigma(\stackrel{\rightarrow}{e_1})\big|
\ -\ \big|\sigma(\stackrel{\rightarrow}{e_2})\big|\ +\ 
\big|\sigma(\stackrel{\rightarrow}{e_1},\stackrel{\rightarrow}{e_2})\big|
=294-56-42+8=204.$$

Dualizing matters in obvious ways one arrives at the same result:
$$ |\rho\cap\sigma|=|\rho|-\big|\rho(\stackrel{\rightarrow}{n_1})\big|
-\big|\rho(\stackrel{\rightarrow}{n_2})\big| -\big|\rho(\stackrel{\rightarrow}{n_3})\big|
\ +\ \big|\rho(\stackrel{\rightarrow}{n_1},\stackrel{\longrightarrow}{n_2})\big|\  
\ +\ \big|\rho(\stackrel{\rightarrow}{n_1},\stackrel{\longrightarrow}{n_3})\big|\  
\ +\ \big|\rho(\stackrel{\rightarrow}{n_2},\stackrel{\longrightarrow}{n_3})\big|\ -\ 
\big|\rho(\stackrel{\rightarrow}{n_1},\stackrel{\rightarrow}{n_2},
         \stackrel{\rightarrow}{n_3})\big|$$
         $$=392-112-56-56+14+16+8-2=204$$

The PIE is easier to grasp than the proof-details of Theorem 1 in Section 9, and also calculating $|\rho\cap\sigma|$ beats anwering
$\rho\cap\sigma\stackrel{?}{=}\es$. Trouble is, PIE takes exponential time $O(m2^m)$, as opposed to $O(m^2)$.
Here $m$ is the number of wildcards in the row which has the fewer wildcards. (In practice time becomes an issue only for $m$ somewhere $>20$, depending on your hardware.)

 {\bf 3.8}  Continuing with the example in 3.7, recall that $|\rho\cap\sigma|=204$. But suppose we need  to list these 204 bitstrings explicitely.
 Here comes how to do it.
 Decompose $\sigma=\sigma_1\uplus\cdots\uplus\sigma_{18}$ into a  disjoint union of 012-rows $\sigma_i$. Then by distributivity $\rho\cap\sigma=
 (\rho\cap\sigma_1)\uplus\cdots\uplus (\rho\cap\sigma_{18})$, and each $\rho\cap\sigma_i$ is  readily expressed as  012e-row:

 \begin{tabular}{l|c|c|c|c|c|c|c|c|c||c|c|c|}
	& 1 & 2 & 3& 4 & 5  & 6   & 7& 8& 9&  &  &   \\ \hline
	& & & & & & & & &  & &\\ \hline
	${\sigma}:=$ &  $n_1$ & $n_1$ & $n_2$ & $n_2$ & $n_2$& $n_3$& $n_3$& $n_3$& $2$ &  & &\\ \hline
      ${\rho}:=$ &  2 & $e_1$ & 2 & $e_1$ & $e_1$& 2 & $e_2$& 
      $e_2$& $e_2$ &  &  &\\ \hline
	& & & & & & & & & & & & $|\tau_i|$\\ \hline\hline
      ${\sigma_1}:=$ & 0 & 2 & 0 & 2 & 2& 0& 2&  2& 2 & 
      $\tau_1:=$ & $0\ e_1\ 0\ e_1\ e_1\ 0\ e_2\ e_2\ e_2$ & 49\\ \hline     
       ${\sigma_2}:=$ & 0 & 2 & 0 & 2 & 2& 1& 0&  2& 2 & 
      $\tau_2:= $& $0\ e_1\ 0\ e_1\ e_1\ 1\ 0\ e_2\ e_2$ & 21\\ \hline
        ${\sigma_3}:=$ & 0 & 2 & 0 & 2 & 2& 1& 1&  0& 2 & 
      $\tau_3:=$&$0\ e_1\ 0\ \ e_1\ e_1\ 1\ 1\ \ 0\ \ 2$ & 14\\ \hline
       ${\sigma_4}:=$ & 0 & 2 & 1 & 0 & 2& 0& 2&  2& 2 & 
      $\tau_4:=$&$0\ e_1\ 1\ 0\ e_1\ 0\ e_2\ e_2\ e_2$ & 21\\ \hline
      
       ${\sigma_5}:=$ & 0 & 2 & 1 & 0 & 2&  1& 0&  2& 2 & 
      $\tau_5:=$&$0\ e_1\ 1\ 0\ e_1\ 1\ 0\ e_2\ e_2$ & 9\\ \hline
      ${\sigma_6}:=$ & 0 & 2 & 1 & 0 & 2&  1& 1&  0& 2 & 
      $\tau_6:=$&$0\ e_1\ 1\ 0\ e_1\  1\ 1\ \ 0\ 2$ & 6\\ \hline
      ${\sigma_7}:=$ & 0 & 2 & 1 & 1 & 0&  0& 2&  2& 2 & 
      $\tau_7:=$&$0\  2\  1\ 1\ 0\ \ 0\ \ e_2\ e_2\ e_2$ & 14\\ \hline
       ${\sigma_8}:=$ & 0 & 2 & 1 & 1 & 0&  1& 0&  2& 2 & 
      $\tau_8:=$&$0\ 2\ 1\ 1\ 0\ 1\ \ 0\ e_2\ e_2$ & 6\\ \hline
       ${\sigma_9}:=$ & 0 & 2 & 1 & 1 & 0&  1& 1&  0& 2 & 
      $\tau_9:=$&$0\ \ 2\ 1\ 1\ \ 0\ 1\ 1\ 0\ 2$ & 4\\ \hline
 
       ${\sigma_{10}}:=$ & 1 & 0 & 0 & 2 & 2& 0& 2&  2& 2 & 
      $\tau_{10}:=$&$1\ 0\ 0\ e_1\ e_1\ 0\ e_2\ e_2\ e_2$ & 21\\ \hline
       ${\sigma_{11}}:=$ & 1 & 0 & 0 & 2 & 2& 1& 0&  2& 2 & 
      $\tau_{11}:=$&$1\ 0\ 0\ e_1\ e_1\ 1\ 0\ e_2\ e_2$ & 9\\ \hline
        ${\sigma_{12}}:=$ & 1 & 0 & 0 & 2 & 2& 1& 1&  0& 2 & 
      $\tau_{12}:=$&$1\ 0\ 0\ e_1\ e_1\ 1\ 1\ 0\ 2$ & 6\\ \hline
       ${\sigma_{13}}:=$ & 1 & 0 & 1 & 0 & 2& 0& 2&  2& 2 & 
      $\tau_{13}:=$&$1\ 0\ 1\ 0\ 1\ 0\ e_2\ e_2\ e_2$ & 7\\ \hline
       ${\sigma_{14}}:=$ & 1 & 0 & 1 & 0 & 2&  1& 0&  2& 2 & 
      $\tau_{14}:=$&$1\ 0\ 1\ 0\ 1\ 1\ 0\ e_2\ e_2$ & 3\\ \hline     
      ${\sigma_{15}}:=$ & 1 & 0 & 1 & 0 & 2&  1& 1&  0& 2 & 
      $\tau_{15}:=$&$1\ 0\ 1\ 0\ 1\ 1\  1 \ \ 0\ 2$ & 2\\ \hline
      ${\sigma_{16}}:=$ & 1 & 0 & 1 & 1 & 0&  0& 2&  2& 2 & 
      $\tau_{16}:=$&$1\ 0\ 1\ 1\ 0\ 0\ e_2\ e_2\ e_2$ & 7\\ \hline
       ${\sigma_{17}}:=$ & 1 & 0 & 1 & 1 & 0&  1& 0&  2& 2 & 
      $\tau_{17}:=$&$1\ 0\ 1\ 1\ 0\ 1\ 0\ e_2\ e_2$ & 3\\ \hline
       ${\sigma_{18}}:=$ & 1 & 0 & 1 & 1 & 0&  1& 1&  0& 2 & 
      $\tau_{18}:=$&$1\ \ 0\ 1\ 1\ \ 0\ 1\ 1\ 0\ 2$ & 2\\ \hline          	
\end{tabular}

{\sl Table 4: Writing $\rho\cap\sigma$ as disjoint union of eighteen 012e-rows }

{\bf 3.8.1} Let us first check that Table 4 achieves what it claims. By inspection  all $\sigma_i$'s are contained in $\sigma$ and their cardinalities sum up to $64+32+\cdots+4+2=294$,
which coincides with the cardinality $3\cdot 7\cdot 7\cdot 2$ of $\sigma$. Therefore $\sigma$ must be the {\it disjoint} union of the $\sigma_i$'s:
$$(4)\quad \sigma=\sigma_1\uplus\cdots\uplus\sigma_{18}$$
For each 012e-row $\tau_i$, as defined on  the right of $\sigma_i$ in Table 4, one readily verifies that 
$$(5)\quad \tau_i\s\rho\cap\sigma_i\ \ for\ all\ \ 1\le i\le 18.$$ 
For instance  $\tau_4=(0,e_1,1,0,e_1,0,e_2,e_2,e_2)$ is evidently contained in $\rho$, and also $\tau_4\s\sigma_4$ because 
$zeros(\tau_4)$  cuts all $n$-wildcards of $\sigma$. The cardinalities of the rows $\tau_i$ sum up to
$49+21+\cdots+3+2=204$, which by 3.7 equals  $|\rho\cap\sigma|$. To summarize,
$$204=|\tau_1|+\cdots+|\tau_{18}|\stackrel{(5)}{\le}|\rho\cap\sigma_1|+\cdots+|\rho\cap\sigma_{18}|
=|\rho\cap(\sigma_1\uplus\cdots\uplus\sigma_{18})|\stackrel{(4)}{=}|\rho\cap\sigma|=204.$$
Therefore $\rho\cap\sigma_i=\tau_i$ and $\rho\cap\sigma=\tau_1\uplus\cdots\uplus\tau_{18}$.

{\bf 3.8.2} So far, so good. But all of this begs two questions. First, how does one generally find 012-rows $\sigma_1,...,\sigma_k$ with $\sigma_1\uplus\cdots\uplus\sigma_k=\sigma$? Second, having found these 012-rows, it can happen (different from above) that some intersections $\rho\cap\sigma_i$ are empty.
If our algorithm aspires to run in output polynomial time, then generating $\sigma_i$'s that yield empty intersections must be prevented. Both problems (the second one being tougher) will be settled in Section 9.

\section{ Closure systems, set-filters and  set-ideals}

It is well known that a CNF of type {\tt Positive$\hspace{2pt}\wedge\hspace{2pt}$Negative} is satisfiable iff a certain set-filter ${\cal F}$ intersects a certain set-ideal ${\cal J}$.
We set out to compute ${\cal J}\cap{\cal F}$ by manipulating 012e- and 012n-rows. Much of Section 4 is either well known or has appeared in the author's previous works. All of this caters for Sections 5,6,7.

{\bf 4.1} For any  set $W$ a set  system $CS\s{\cal P}(W)$ is called a {\it closure system} if $W\in CS$ and if $X\cap Y\in CS$ for all $X,Y\in CS$. Here comes one way to capture closure systems. 
An {\it implication} on a set $W$ is an ordered pair $(A,B)\in{\cal P}(W)\times{\cal P}(W)$.
We henceforth write $A\ra B$ instead of $(A,B)$, and call $A$ the {\it premise}, and $B$ the {\it conclusion} of the implication.
A set $X\s W$ is said to {\it satisfy} $A\ra B$ if either $A\not\s X$ or $B\s X$. In the first case the satisfaction occurs
{\it premise-wise}, in the second case {\it conclusion-wise}. (Both can happen simultaneously.) Let $\Sigma$ be a family of implications on $W$. One says that $X\s W$ is {\it $\Sigma$-closed } if $X$ satisfies all implications in $\Sigma$. As is well known, the family $CS(\Sigma)$ of all $\Sigma$-closed sets is a closure system.

A set system ${\cal J}\s{\cal P}(W)$ is  a {\it set-ideal} if  for all $Z\in{\cal J}$ it follows from $Y\s Z$ that $Y\in{\cal J}$.  Here comes one way how set-ideals arise. Let $\HH \s{\cal P}(W)$ be any set system (often called {\it hypergraph}). Then a set $X\s W$ is a {\it noncover} (w.r.t. $\HH $) if $X\not\supseteq H$ for all $H\in\HH $. Clearly the family $NC(\HH )$ of all noncovers is a set-ideal. 

Dually, a set system ${\cal F}\s{\cal P}(W)$ is  a {\it set-filter} if  for all $Z\in{\cal F}$ it follows from $Z\s Y$ that $Y\in{\cal F}$. One source of set-filters is this. For any hypergraph $\HH \s{\cal P}(W)$ one calls $Y\s W$  a {\it hitting set (or: transversal)} of  $\HH $ if $Y\cap H\neq\es$ for all $H\in\HH $. Clearly the family $HS(\HH )$ of all hitting sets is a set-filter.

We next survey algorithms to calculate closure systems (4.2), set-ideals (4.3), and set-filters (4.4).

{\bf 4.2} Consider $W:=\{1,2,...,5\}$ and the family of implications\\ $\Sigma:=\big\{\ \{1,2,3\}\ra\{5\},\ \{4,5\}\ra\{1\},\ \{3,4\}\ra\{2\}\ \big\}$.
Upon identifying subsets with bitstrings  it holds that $CS(\Sigma)=Mod(f_1)$, where
$$(6)\quad f_1:=(\ol{x}_1\vee\ol{x}_2\vee\ol{x}_3\vee x_5)\wedge (\ol{x}_4\vee\ol{x}_5\vee x_1)\wedge (\ol{x}_3\vee\ol{x}_4\vee x_2)$$
is\footnote{It has the extra property that all Horn-clauses are {\it pure}, i.e. contain a positive literal.} a {\tt Horn-CNF}. 
Here comes a sketch of how the {\it implication n-algorithm} in [W2] "imposes" the implications in $\Sigma$ one by one
(in the given order). For starters, the 2n-row $s_0$ in Table 5 contains exactly those $X\in{\cal P}(W)$ which premise-wise satisfy $\{1,2,3\}\ra\{5\}$. The sets $X\in{\cal P}(W)$ that conclusion-wise,  but {\it not} premise-wise, satisfy $\{1,2,3\}\ra\{5\}$, are the members of $s_1$.
Incidentally $s_1$ happens to be {\it final}, in the sense that all $X\in s_1$ are $\Sigma$-closed; indeed, $X$ satisfies $\{4,5\}\ra\{1\}$ and  $\{3,4\}\ra\{2\}$ conclusion-wise since $1,2\in X$. 
Hence $s_1$ is removed  and stored in a safe place.
As to $s_0$, e.g. $\{2,3,4,5\}\in s_0$ does not satisfy $\{4,5\}\ra\{1\}$, and so this implication is {\it pending} to be imposed upon $s_0$. 

To do so, clearly $s_{00}$ is the set of all
$X\in s_0$ that premise-wise satisfy $\{4,5\}\ra\{1\}$, and $s_{01}$ is the set of all
$X\in s_0\setminus s_{00}$ that conclusion-wise satisfy $\{4,5\}\ra\{1\}$. In both these rows the pending implication is $\{3,4\}\ra\{2\}$. Turning to $s_{00}$ first\footnote{In fact, what we employ is a Last-In-First-Out (LIFO) stack. This is a fundamental data structure in computer science which is taylored for depth first search.}, we leave it to the reader to verify that $s_{000}\uplus s_{000}'$ is the family of those $X\in s_{00}$ that
premise-wise satisfy $\{3,4\}\ra\{2\}$ (take note of the bold Abraham 0-Flag as defined in Section 2). And $s_{001}$ collects the sets $X\in s_{00}$ that satisfy $\{3,4\}\ra\{2\}$ conclusion-wise. 
Upon storing the three final rows the only row in our LIFO stack is $s_{01}$.
There is no $X\in s_{01}$ that satisfies $\{3,4\}\ra \{2\}$ conclusion-wise (why?), and so $s_{01}$ gives rise to the final row $s_{010}$. Removing it, the LIFO stack becomes empty, and the algorithm stops.  We found  that $|CS(\Sigma)|=|s_0|+\cdots+|s_{010}|=2+12+6+1+2=23$.

The implication $n$-algorithm works in polynomial total time, i.e. $O(Rh^2w^2)$ where $w:=|W|,\ h:=|\HH|,$ and $R$ is the number of output 012n-rows. Essential in the proof is the concept of
feasibility. Specifically, e.g. in Table 5 the rows $s_{00}$ and $s_{01}$ are the {\it candidate sons} of row $s_0$. Generally a candidate row is {\it feasible} if it contains at least one model (in our case: at least one $\Sigma$-closed set). Infeasible rows must be cancelled because otherwise an output polynomial running time of the overall algorithm can hardly be established. It hence is crucial to have polynomial time feasibility test; see [W2] for more details.

\begin{tabular}{l|c|c|c|c|c|l}
	& 1 & 2 & 3& 4 & 5     & \\ \hline
	& & & & & &   \\ \hline
	${s_0}:=$ &  $n$ & $n$ & $n$ & $2$ & $2$& pending $\{4,5\}\ra\{1\}$  \\ \hline
    ${s_1}:=$ &  $1$ & $1$ & $1$ & $2$ & $1$&  final \\ \hline
    & & & & & &   \\ \hline
	${s_0}:=$ &  $n$ & $n$ & $n$ & $2$ & $2$& pending $\{4,5\}\ra\{1\}$  \\ \hline
    & & & & & &   \\ \hline
    $s_{00}:=$ &  $n_1$ & $n_1$ & $n_1$ & $n_2$ & $n_2$& pending $\{3,4\}\ra\{2\}$  \\ \hline
     $s_{01}:=$ &  $1$ & $n_1$ & $n_1$ & $1$ & $1$&  pending $\{3,4\}\ra\{2\}$ \\ \hline
    & & & & & &   \\ \hline
     $s_{00{\bf 0}}:=$ &  $2$ & $2$ & {\bf 0} & {$\bf n_2$} & $n_2$&  final \\ \hline
      $s'_{00{\bf 0}}:=$ &  $n_1$ & $n_1$ & {\bf 1} & {\bf 0} & $2$& final  \\ \hline
       $s_{00{\bf 1}}:=$ &  $0$ & $1$ & $1$ & $1$ & $0$& final  \\ \hline
        $s_{01}:=$ &  $1$ & $n_1$ & $n_1$ & $1$ & $1$&  pending $\{3,4\}\ra\{2\}$ \\ \hline
         & & & & & &   \\ \hline
          $s_{01}:=$ &  $1$ & $n_1$ & $n_1$ & $1$ & $1$&  pending $\{3,4\}\ra\{2\}$ \\ \hline
         & & & & & &   \\ \hline
          $s_{010}:=$ &  $1$ & $2$ & $0$ & $1$ & $1$&  final \\ \hline
    \end{tabular}

    {\sl Table 5: Writing $Mod(f_1)$ as disjoint union of five 012n-rows.}

{\bf 4.3}  If we drop the positive literals in $(6)$ we get the {\tt Negative-CNF}
$$(7)\quad f_2:=(\ol{x}_1\vee\ol{x}_2\vee\ol{x}_3)\wedge (\ol{x}_4\vee\ol{x}_5)\wedge (\ol{x}_3\vee\ol{x}_4).$$
The models of $f_2$ clearly match the  noncovers  of the hypergraph $\HH (f_2):=\{\{1,2,3\},\{4,5\},\{3,4\}\}$.
In other words, $Mod(f_2)$  coincides with ${\cal J}(f_2):=NC(\HH (f_2))$. The {\it noncover n-algorithm} (briefly: {\it n-algorithm}), which is an obvious simplification of the implication $n$-algorithm, represents ${\cal J}(f_2)$ as a {\it disjoint} union of 012n-rows (Table 6). In particular
$|{\cal J}(f_2)|=12+6=18$.

\begin{tabular}{l|c|c|c|c|c|l}
	& 1 & 2 & 3& 4 & 5     & \\ \hline
	& & & & & &   \\ \hline
	$\sigma:=$ &  $n_1$ & $n_1$ & $n_1$ & $n_2$ & $n_2$& pending $\{3,4\}$  \\ \hline
    & & & & & &   \\ \hline
	$\sigma_0:=$ &  $2$ & $2$ & {\bf 0} & $\bf n_2$ & $n_2$&  final \\ \hline
    $\sigma_1:=$ &  $n_1$ & $n_1$ & {\bf 1} & $\bf 0$ & $2$&  final \\ \hline
    \end{tabular}

{\sl Table 6: The $n$-algorithm writes the set-ideal  ${\cal J}(f_2)=Mod(f_2)$ as $\sigma_0\uplus\sigma_1$.}

{\bf 4.4} Upon switching all literals from negative to positive, $(7)$ becomes
$$(8)\quad g_2:=({x}_1\vee{x}_2\vee{x}_3)\wedge ({x}_4\vee{x}_5)\wedge ({x}_3\vee{x}_4)$$
The models of $g_2$ are now the hitting sets of the hypergraph $\HH (g_2):=\{\{1,2,3\},\{4,5\},\{3,4\}\}$ (which happens to be $\HH (f_2)$).  Thus $Mod(g_2)$ coincides with ${\cal F}(g_2):=HS(\HH (g_2))$.
The {\it transversal\footnote{We keep the terminology of previous publications; recall that "transversal" is just another name for "hitting set".} e-algorithm} (briefly: {\it e-algorithm}) is a mirror image of the noncover n-algorithm:

\begin{tabular}{l|c|c|c|c|c|l}
	& 1 & 2 & 3& 4 & 5     & \\ \hline
	& & & & & &   \\ \hline
	$\rho:=$ &  $e_1$ & $e_1$ & $e_1$ & $e_2$ & $e_2$& pending $\{3,4\}$  \\ \hline
    & & & & & &   \\ \hline
	$\rho_0:=$ &  $2$ & $2$ & {\bf 1} & $\bf e_2$ & $e_2$&  final \\ \hline
    $\rho_1:=$ &  $e_1$ & $e_1$ & {\bf 0} & $\bf 1$ & $2$&  final \\ \hline
    \end{tabular}

{\sl Table 7: The $e$-algorithm writes the set-filter ${\cal F}(g_2)=Mod(g_2)$ as $\rho_0\uplus\rho_1$.}

Is it a coincidence that $|{\cal F}(g_2)|=18=|{\cal J}(f_2)|$? No, because for any hypergraph $\HH \s{\cal P}(W)$ it holds (putting $Y^c:=W\setminus Y$) that 
$$(9)\quad HS(\HH )=\{Y^c:\ Y\in NC(\HH )\}\quad and\quad NC(\HH )=\{X^c:\ X\in HS(\HH )\}$$

In the remainder of Section 4 we are mostly concerned with moving from the individual set-systems
 ${\cal J}, {\cal F}$ towards  ${\cal J}\cap {\cal F}$.
 
{\bf 4.5}  Let $f:\{0,1\}^m\to\{0,1\}$ be any Boolean function defined by a {\tt Negative-CNF} and let ${\cal J}(f)\s{\cal P}(\{1,..,m\})$ be the coupled set-ideal. Likewise let $g:\{0,1\}^m\to\{0,1\}$ be any Boolean function defined by a {\tt Positive-CNF} and let ${\cal F}(g)\s{\cal P}$ be the coupled set-filter. Hence $f\wedge g$ is of type {\tt Negative$\hspace{2pt} \wedge\hspace{2pt} $Positive} and it holds that $Mod(f\wedge g)={\cal J}(f)\cap{\cal F}(g)$.
Upon expressing the set-filter ${\cal F}(g)$ as disjoint union of 012e-rows $\rho_i$, and the set-ideal ${\cal J}(f)$ as disjoint union of 012n-rows $\sigma_j$ (as glimpsed in 4.3 and 4.4), it holds that
$$(10)\quad {\cal J}(f)\cap{\cal F}(g)\neq \es\ \LRa\ (\exists i)(\exists j)\ \rho_i\cap\sigma_j\neq\es$$
Several problems arise. {\it Problem 1}, how fast can the emptiness of {\it one set} $\rho_i\cap\sigma_j$ be decided?
{\it Problem 2}, how many intersections $\rho_i\cap\sigma_j$ need to be evaluated? {\it Problem 3}, what if
$ {\cal J}(f)\cap{\cal F}(g)\neq \es$ has been settled but all members of $ {\cal J}(f)\cap{\cal F}(g)$ need to be known?
 {\it Problem 4}, how to handle set-ideals $\cal J$ and set-filters  ${\cal F}$ which are not initially given in the form
 ${\cal J}= {\cal J}(f)$ and ${\cal F}= {\cal F}(g)$?

 {\bf 4.6} Let us convey, to some extent, the solutions of these problems.  {\it Problem 1} is fully settled by Theorem 1.
 As to  {\it Problem 2} (the most vexing one), there can indeed be plenty ($=:N$) intersections $\rho_i\cap\sigma_j$ to be evaluated. A case in point is Section 6 where hopefully symmetry exploitation will eventually shrink $N$.
 As to  {\it Problem 3}, in Section 9 we show how to expand nonempty sets $\rho_i\cap\sigma_j$ fast into disjoint unions of 012e-rows. Consequently also $ {\cal J}(f)\cap{\cal F}(g)$ can be expanded in this way. Concerning {\it Problem 4}, more comments follow in 4.6.1 to 4.6.4.

{\bf 4.6.1}  If the set-ideal ${\cal J}\s{\cal P}(W)$ is not given as ${\cal J}= {\cal J}(f)$  for some {\tt Negative-CNF} $f$, then ${\cal J}$ is likely given (we omit other possibilities) by its
 inclusion-maximal members (called {\it facets}) $F_1,...,F_t\in {\cal J}$. They uniquely determine ${\cal J}$ in that
${\cal J}={\cal P}(F_1)\cup{\cal P}(F_2)\cup\cdots\cup{\cal P}(F_t)$. Let us address the inconvenience that this union is never\footnote{Because $\es\in{\cal P}(F_i)$ for all $1\le i\le t$.} disjoint.
 If say $W:=\{1,...,m\}$, let $r_i$ be the 02-row of length $m$ defined by $twos(r_i):=F_i, \ zeros(r_i):=W\setminus F_i$.
It then holds that ${\cal J}=r_1\cup\cdots\cup r_t$.  Upon applying the "detachment method" of Section 2 one obtains
${\cal J}=\rho'_1\uplus\cdots\uplus \rho'_p$ for certain 012-rows $\rho'_i$. (As shown in 4.6.3, an extra effort yields 012e-rows $\rho_i'$.)

{\bf 4.6.2} Dually, if the set-filter ${\cal F}\s{\cal P}(W)$ is not given as ${\cal F}= {\cal F}(g)$  for some {\tt Positive-CNF} $g$, then ${\cal F}$ is likely given by its inclusion-minimal members (called {\it generators}) $G_1,...,G_s\in {\cal F}$.
They uniquely determine ${\cal F}$ in that
${\cal F}=G_1\!\uparrow\cup G_2\!\uparrow\cup  \cdots\cup G_s\!\uparrow$. (For any $G\s W$ put  $G\!\uparrow\ :=\{X\in{\cal P}(W):\ X\supseteq G\}$ .)
Let us fix again the fact that this union is never disjoint.
If again $W=\{1,...,m\}$, let $r_i'$ be the 12-row of length $m$ defined by $ones(r_i'):=G_i,\ twos(r_i'):=W\setminus G_i$. It then holds that ${\cal F}=r_1'\cup\cdots\cup r_s'$. 
Upon applying Section 2 one obtains ${\cal F}=\sigma_1'\uplus\cdots\sigma_q'$ for certain 012-rows $\sigma_i'$.
(As shown in 4.6.3, an extra effort yields 012n-rows $\rho_i'$.)

{\bf 4.6.3} First off, this somewhat technical Subsection can be skipped without loosing the story line. For the sake of simplicity the example in Section 2 was taylored to the effect that only one kind of "detachment" occured (which triggered Abraham 1-Flags). Let us step that up to Abraham 0-Flags and Abraham 01-Flags! Consider thus the 012-rows $r_1,\ r_2$ in Table 8. An argument dual to the one in Section 2 (switching 0's and 1's) shows that
$r_1\setminus r_2=r_3\uplus r_4\uplus r_5$, and all is based on some (boldface) $3\times 3$ Abraham 0-Flag. Next,
$\ol{r_2}$ arises from $r_2$ by turning two 2's to 0's. Looking at
 $r_1\setminus\ol{r_2}$, now somehow {\it both} kinds of Abraham Flags seem to be required. This is true; specifically a $s\times s$ Abraham 0-Flag $A_0$ and a $t\times t$ Abraham 1-Flag $A_1$ are compiled to some $(s+t)\times (s+t)$ Abraham 01-Flag.  The case $s=3,\ t=2$ is illustrated  in Table 8, i.e. $r_1\setminus\ol{r_2}=r_6\uplus\cdots\uplus r_{10}$.

\begin{tabular}{l|c c c|c c|c c c|l}
\hline
	${r_1}:=$ &  2 & 2 & 2 & 2 & 2& 0 & 1 &1 \\ \hline
    ${r_2}:=$ & 1 & 1 & 1 & 2 & 2& 2 &1&  2 \\ \hline
     $\ol{r_2}:=$ & 1 & 1 & 1 &\bf  0 &\bf 0& 2 &1 &  2 \\ \hline
    & & & & & & & &  \\ \hline
    ${r_3}:=$ & \bf 0 & \bf 2 & \bf 2 & 2 & 2& 0 &1 &  1 \\ 
     ${r_4}:=$ & \bf 1 & \bf 0 & \bf 2 & 2 & 2& 0 &1 &  1 \\ 
      ${r_5}:=$ & \bf 1 & \bf 1 & \bf 0 & 2 & 2& 0 &1 &  1 \\ \hline
       & & & & & & & &  \\ \hline
       ${r_6}:=$ & \bf 0 & \bf 2 & \bf 2 & 2 & 2& 0 &1 &  1 \\ 
     ${r_7}:=$ & \bf 1 & \bf 0 & \bf 2 & 2 & 2& 0 &1 &  1 \\ 
      ${r_8}:=$ & \bf 1 & \bf 1 & \bf 0 & 2 & 2& 0 &1 &  1 \\ \hline
      ${r_9}:=$ &  1 & 1 & 1 & \bf 1 &\bf 2& 0 &1 &  1 \\ 
       ${r_{10}}:=$ &  1 & 1 & 1 & \bf 0 &\bf 1& 0 &1 &  1 \\ \hline
        & & & & & & & &  \\ \hline
         ${r_{11}}:=$ & \bf n & \bf n & \bf n & 2 & 2& 0 &1 &  1 \\
          ${r_{12}}:=$ &  1 & 1 & 1 & \bf e &\bf e& 0 &1 &  1 \\ \hline
       
    \end{tabular}

    {\sl Table 8: Introducing Abraham 01-Flags}

If one embraces n-wildcards and e-wildcards, then  each $(s+t)\times (s+t)$ Abraham 01-Flag  can be replaced by some Abraham $ne$-Flag with just {\it two} rows, as shown in Table 8. Check that indeed $|r_6\uplus\cdots \uplus r_{10}|=31=|r_{11}\uplus r_{12}|$. This may look like a brilliant shortcut. The flipside is that upon iteration detachment has to be achieved w.r.t. 012ne-rows, and this gets convoluted. Suffice it to say that the task is somewhat easier if the initial  012-rows are all 02-rows (= the facets $F_i$), or are all 12-rows (=the generators $G_i$). In these cases the output (disjoint) rows are all of type 012e, respectively type 012n. The former case has been dealt with in detail in [W3].

{\bf 4.6.4} If $\cal J$ is given by its facets $F_1,...,F_t$, and $\cal F$ is given by its generators $G_1,...,G_s$, and the question is merely ${\cal J}\cap{\cal F}\stackrel{?}{=}\es$, then this can be settled at once (notice the relation to (10)):
$$(10')\quad {\cal J}\cap{\cal F}\neq \es\ \LRa\ (\exists i)(\exists j)\ F_j\s G_i.$$
Accordingly the next concept fits in well. A set-family
(possibly empty) $Co\s{\cal P}(W)$ is called {\it convex} if for all $X,Y\in Co$
and all $Z\s W$ it follows from $X\s Z\s Y$ that $Z\in Co$. For instance, if $\cal J$ is a set-ideal and $\cal F$ a set-filter, then ${\cal J}\cap{\cal F}$ is convex. Conversely, {\it every} convex set-family $Co\s{\cal P}(W)$
can be written as $Co={\cal J}\cap{\cal F}$. Indeed, the minimal members of $Co$ yield the generators of $\cal F$, and the maximal members of $Co$ yield the facets  of $\cal J$.

\section{ Compressing the family of all minimal hitting sets\\ of a hypergraph}

Recall from Subsection 4.6.2, if a set-filter\footnote{Set-ideals behave dually to the forthcoming.} ${\cal F}\s{\cal P}(W)$ is given by its generators, then it can be represented by 012n-rows. And if  ${\cal F}$ is given by a positive CNF $g$, it can be represented by 012e-rows.
Recall that $Mod(g)=HS(\HH )$, where $\HH$ is the hypergraph whose hyperedges match the clauses of $g$. Consider now this related problem: 

{\it Find the {\bf generators} of  $\cal F$ when it is given as  ${\cal F}:=HS(\HH )$. In other words, find the family $MHS(\HH )$ of all {\bf minimal} hitting sets!}

It is a notorious open question whether $MHS(\HH )$ can be enumerated in polynomial total time; we recommend the survey [GV].
In 5.1 we show how to find the better behaved subfamily of all {\bf minimum} (cardinality) hitting sets, and in 5.2  glimpse at the general case. 

{\bf 5.1} For any set system $\cal S$ we write $Mmal({\cal S})$ for the subfamily of all minimal members of $\cal S$, and $Mmum({\cal S})\s Mmal({\cal S})$ for the family of all minimum members. 
Consider the hypergraph
$\HH_0:=\{\{1,2,4,5\},\{1,3,6\},\{2,7\}\}\s {\cal P}(\{1,...,7\})$. Applying the $e$-algorithm of 4.4 yields $HS(\HH_0)=\rho_1\uplus\rho_2\uplus\rho_3$, where

\begin{tabular}{l|c|c|c|c|c|c|c|c}
	& 1 & 2 & 3& 4 & 5  & 6   & 7 &\\ \hline
	& & & & & & & & \\ \hline
	${\rho_1}:=$ &  $0$ & $1$ & $e$ & $2$ & $2$& $e$& $2$& $\ni 23, 26$  \\ \hline  
    ${\rho_2}:=$ &  $0$ & $0$ & $e'$ & $e$ & $e$& $e'$& $1$ & $\ni 734, 735, 764, 765$  \\ \hline
    ${\rho_3}:=$ &  $1$ & $e$ & $2$ & $2$ & $2$& $2$& $e$  & $\ni 12, 17$ \\ \hline
    & & & & & & & & \\ \hline
	${\sigma_1}:=$ &  $0$ & $0$ & $n'$ & $n$ & $n$& $n'$& $2$ &  \\ \hline
    ${\sigma_2}:=$ &  $0$ & $1$ & $n$ & $0$ & $0$& $n$& $0$ &   \\ \hline
    ${\sigma_3}:=$ &  $1$ & $n$ & $0$ & $0$ & $0$& $0$& $n$ &  \\ \hline
\end{tabular}

{\sl Table 9:  $HS(\HH_0)=\rho_1\uplus\rho_2\uplus\rho_3$ and 
$MC(\HH_0)=\sigma_1\uplus\sigma_2\uplus\sigma_3$ }

For any 012e-row $\rho$, viewed as set system, let us calculate $Mmal(\rho)$ and $Mmum(\rho)$. To begin with, $X\in\rho$ belongs to $Mmal(\rho)$ iff $X$ is the union of $ones(\rho)$ with some transversal of  $\{pos(e_1),pos(e_2),..\}$.
Consider say $\rho=\rho_2$. The transversals of $\{pos(e'), pos(e)\}=\{\{3,6\},\{4,5\}\}$ are (using shorthand notation)
$34,35,64,65$. In view of $ones(\rho_2)=\{7\}$ one gets $Mmal(\rho_2)=\{734,735,764,765\}$. Likewise 
$Mmal(\rho_1)=\{26,23\}$ and $Mmal(\rho_3)=\{12,17\}$. It is now clear that for {\it each} 012e-row $\rho$ (not just in Table 9) it holds that
$$Mmum(\rho)=Mmal(\rho)$$

{\bf 5.1.1} Recall that generally only $Mmum({\cal S})\s Mmal({\cal S})$. 
Let us have a closer look at the scenario ${\cal S}:=HS(\HH )$.
For any hypergraph $\HH$ we put $\mu(\HH ):=min\{|X|:\ X\in HS(\HH )\}$. Thus $\mu(\HH )$ is the common  cardinality of all $X$ in $ MCHS(\HH ):=Mmum(HS(\HH ))$. Here MCHS stands for "minimum cardinality hitting sets". Suppose that
$$(11)\quad HS(\HH )=\rho_1\uplus\rho_2\uplus\cdots\uplus\rho_t\ for\ disjoint\ 012e\!-\!rows\ \rho_i$$
 Take any $X\in MCHS(\HH )$. A fortiori $X\in Mmum(\rho_j)$, where $\rho_k$ is the unique row that contains $X$. It follows that $\mu(\rho_k)=\mu(\HH )$ and that 
$Mmum(\rho_k)\s MCHS(\HH )$. Therefore:
\begin{itemize}
    \item[(12)]{\it  $MCHS(\HH )$ is the disjoint union of those $Mmum(\rho_k)$ where $\rho_k$ in (11) happens\\ to satisfy
    $\mu(\rho_k)=\mu(\HH ).$}
\end{itemize}

But how to calculate $\mu(\HH )$ in the first place? For some\footnote{For instance, if $\HH$ is the family of all connected edge sets of a connected graph on $n$ vertices, then $\mu(\HH )=n-1$.} hypergraphs $\HH$ one may know $\mu(\HH )$
in advance. Otherwise $\mu(\HH )$ is easily obtained, {\it provided} the representation (11) has been achieved, namely
$\mu(\HH )=min\{\mu(\rho_1),\mu(\rho_2),...,\mu(\rho_t\}$.
Thus, having computed $\rho_1,\rho_2,\rho_3$ in Table 9, one reads off that
$\mu(\HH _0)=min\{\mu(\rho_1),\mu(\rho_2),\mu(\rho_3)\}=2$, and so (12) implies that
$$MCHS(\HH _0)=Mmum(\rho_1)\uplus Mmum(\rho_3)=\{23,26\}\uplus\{12,17\}$$

  {\bf 5.1.2} The {\it g-wildcard} $(g,g,..,g)$ was useful in previous articles, and will be so in the present article (here and in Section 9). Loosely speaking it signifies "exactly one 1-bit in this area". For\footnote{Of course $(g,g,g)\s (e,e,e))$.} instance $(g,g,g):=\{(1,0,0),(0,1,0),(0,0,1)\}$. We define 012g-rows akin to 012e-rows and 012n-rows. This e.g. yields the compact representation
 $$MCHS({\HH_0})=(0,1,g,2,2,g,2)\uplus (1,g,2,2,2,2,g).$$

\vspace{3mm}

{\bf 5.2} We now turn from $MCHS(\HH )$ to the more cumbersome family $MHS(\HH )$. Each $Y\in MHS(\HH )$ a fortiori belongs to $Mmal(\rho_k)$ where $\rho_k$ is the unique 012e-row in (11) that contains $Y$.
Consequently
$$(13)\quad MHS(\HH )\s Mmal(\rho_1)\uplus\cdots\uplus Mmal(\rho_t) $$
Incidently (by inspection) $\s$ becomes $=$ for $\HH:=\HH_0$. To spell it out:
$$(14)\quad MHS(\HH_0)=\{23,26\}\uplus\{734,735,764,765\}\uplus\{12,17\}$$

In general the inclusion $\s$ in (13) is proper. In fact, some rows $\rho_j$ may be "bad" [W4,p.10] in the sense that $Mmal(\rho_j)\cap MHS(\HH)=\es$.
One way (among several others in [W4]) to sieve $MHS(\HH )$ from a type (13) representation of $HS(\HH )$ proceeds as follows. By definition $X\in {\cal P}(W)$ belongs to $MC(\HH )$ iff each $a\in X$ has at least one  {\it private} hyperedge $H\in \HH $  in the sense that $X\cap H=\{a\}$ (as opposed to merely $\supseteq$). The acronym MC  refers to some kind of "minimal condition" in [MU]. See [W4, Thm.2] for a fresh proof of the  fact (established in [MU] with other terminology) that 
$$(15)\quad MHS(\HH)=HS(\HH )\cap MC(\HH )$$

Since obviously $MC(\HH)$ is a set-ideal, it follows from (15) that $MSH(\HH)$ is a set-system of type ${\cal J}(f)\cap{\cal F}(g)$, i.e. the modelset of some {\tt Negative$\wedge$Positive} type CNF $f\wedge g$.
It is shown in [W4] how $MC(\HH )$ can be represented as a disjoint union of 012n-rows (finding $f$ is subtle). For instance
$MC(\HH_0)=\sigma_1\uplus\sigma_2\uplus\sigma_3$, where the latter rows are defined in Table 9.
In tandem with (11) and (15) follows that $MHS({\HH_0})$ is the disjoint union of all nine sets $\rho_i\cap \sigma_j$. One verifies ad hoc that
$$\rho_1\cap\sigma_2=\{\{2,3\},\{2,6\}\},\ \rho_3\cap\sigma_3=\{\{1,2\},\{1,7\}\},\ \rho_2\cap\sigma_1=\{\{3,4,7\},\{3,5,7\},\{4,6,7\},\{5,6,7\}\},$$
and that otherwise $\rho_i\cap\sigma_j=\es$. This matches (14).

\section{About Ramsey numbers}

Also this Section is about {\tt Negative$\wedge$Positive}. Thus
let us lay out how 012e- and 012n-rows relate to Ramsey numbers. Recall that for any fixed integers $k,\ell\ge 1$ there is an integer (the {\it Ramsey number})  $R(k,\ell)$ with these properties:
\begin{itemize}
    \item If $G=(V,E)$ is any graph with $|V|\ge R(k,\ell)$ then $G$ has a $k$-clique or an $\ell$-anticlique (or both).
    \item $R(k,l)$ is the smallest integer with the stated property. Therefore: For each $m<R(k,\ell)$ there are counterexamples $G_0=(V_0,E_0)$, i.e. $|V_0|=m$ and $G_0$ neither has $k$-cliques nor $\ell$-anticliques.
\end{itemize}

In order to connect the matter to set-ideals and set-filters we put $V_m:=\{1,2,..,m\}$ and focus on  the ${m}\choose{2}$-element set $E_m$ of all edges of the complete graph $K_m:=(V_m,E_m)$. Each $X\in {\cal P}(E_m)$ yields the spanning\footnote{Recall that a subgraph $H$ of a graph $G$ is {\it spanning} if $H$ has the same vertex-set as $G$.} subgraph $(V_m,X)$ of  $K_m$, and each spanning subgraph arises this way. Thinking about $V_m$ in the back of our minds we henceforth identify ${\cal P}(E_m)$ with
the family of all spanning subgraphs of $K_m$.

{\bf 6.1} Let ${\cal J}(k,m)\s {\cal P}(E_m)$ be the set of those spanning subgraphs that have no, i.e. {\it avoid all}, $k$-element cliques of $K_m$. Evidently ${\cal J}(k,m)$ is a set-ideal in $ {\cal P}(E_m)$.
Dually let  ${\cal F}(\ell,m)\s {\cal P}(E_m)$ be the set of those spanning subgraphs of $K_m$ that  avoid all  $\ell$-element anticliques. Evidently ${\cal F}(\ell,m)$ is a set-filter in $ {\cal P}(E_m)$. Take a moment to convince yourself:
$$(16)\quad m\ge R(k,\ell)\ \LRa\  {\cal J}(k,m)\cap {\cal F}(\ell,m)=\es.$$
Property (16) shows that calculating Ramsey numbers can in principle be reduced  to (repeatedly) deciding whether certain set-ideals intersect certain set-filters. In the remainder of Section 6 we ponder how to bring "in principle" closer to "in practice".

{\bf 6.2} In order to add 012n-rows and 012e-rows to the picture, recall from 4.3 and 4.4 that applying the $n$-algorithm to any hypergraph $\HH$ represents $NC(\HH )$ as disjoint union of 012n-rows, and
applying the e-algorithm to any hypergraph $\HH'$ represents $HS(\HH')$ as disjoint union of 012e-rows. Let
$\HH_{k,m}$ be the hypergraph of all $k$-cliques of $K_m$. (For instance, $\{47,49,79\}\in \HH_{3,m}$ for all $m\ge 9$.) Therefore ${\cal J}(k,m)=NC(\HH_{k,m})$ can be displayed via 012n-rows $\sigma_j$. 

What about ${\cal F}(\ell,m)$? First observe the following: Some spanning subgraph $(V_m,X)$ has some fixed $\ell$-anticlique $\{a,b,c,..\}\s V_m$ (thus no edges among any of the vertices $a,b,c,..$) iff the spanning subgraph
$(V_m,X^c)$ has the corresponding $\ell$-clique (thus $\{ab,ac,bc,..\}\s X^c$). Consequently:
\begin{itemize}
\item[${\cal F}(\ell,m)$] $=\{Y^c\in{\cal P}(E_m):\ Y^c\  avoids\  all\ \ell\!-\!anticliques\ 
of\ K_m\} 
$ 
\item[] $=\{Y^c\in{\cal P}(E_m):\ Y\  avoids\  all\ \ell\!-\!cliques\ of\ K_m\}$

\item[]$=\Big\{Y^c\in{\cal P}(E_m):\ Y\in NC({\HH}_{\ell,m})\}\stackrel{(9)}{=} HS({\HH}_{\ell,m})$
\end{itemize}

Suppose ${\cal F}(\ell,m)=HS({\HH}_{\ell,m})$ is displayed via disjoint 012e-rows $\rho_i$ It therefore follows from (16) that
$$(17)\quad m<R(k,\ell)\ \LRa\ (\exists i)(\exists j)\ \rho_i\cap\sigma_j\neq\es$$

{\bf 6.3} To fix ideas, it takes 29  (012e)-rows $\rho_i$ to represent ${\cal F}(3,5)$ and (by symmetry) also 29 012n-rows $\sigma_j$ to represent ${\cal J}(3,5)$. Hence a hefty $29^2=841$ pairs  $(\rho_i,\sigma_j)$ arise in (17). The good news is this. Suppose we strive to show that $5<R(3,3)$. Then, by (17), finding just {\it one pair} $(\rho_i,\sigma_j)$ with $\rho_i\cap\sigma_j\neq\es$ achieves our goal.
Upon trail and error one may find $\sigma=\sigma_j$ and $\rho=\rho_i$ in Table 10.

\begin{tabular}{l|c|c|c|c|c|c|c|c|c|c|l}
	& 12 & 13 & 14& 15 & 23  & 24   & 25& 34& 35& 45 & \\ \hline
	& & & & & & & & & & & \\ \hline
	${\sigma}:=$ &  1 & 1 & 0 & 0 & 0& $n'$& $n'$&  $n$& $n$ & 1 \\ \hline
    ${\rho}:=$ &  1 & 1 & 0 & 0 & 0& $e$& $e'$&  $e$& $e'$ & 1 \\ \hline
    \end{tabular}

    {\sl Table 10: Fact (17) in the context of $m=5$ and $\ell=k=3$}

As we know from 3.4, it follows from $ones(\rho)=ones(\sigma)$ and $zeros(\rho)=zeros(\sigma)$ that
$\rho\cap\sigma\neq\es$. A concrete member of $\rho\cap\sigma$ is $\{13,34,45,25,21\}$, i.e. the pentagon
$1-3-4-5-2-1$ which indeed avoids all  3-cliques and all 3-anticliques (make a sketch).

{\bf 6.4} Here we try to repeat for $K_6$ the success we had with $K_5$ in 6.3. 
For $m\ge 6$ "trial and error" won't work for $K_m$. Instead
we start with the 012e-row 
$\tilde{\rho}\s {\cal F}(3,6)$ (check the inclusion), and try to construct some 012n-row $\tilde{\sigma}$ that
avoids both the ${{6}\choose{3}}=20$ size 3 anticliques, as well as all trivial reasons for $\tilde{\rho}\cap\tilde{\sigma}=\es$.
By Theorem 1 this guarantees $\tilde{\rho}\cap\tilde{\sigma}\neq\es$.

\begin{tabular}{l|c|c|c|c|c|c|c|c|c|c|c|c|c|c|c|c}
	& 12 & 13 & 14& 15 & 16  & 23   & 24& 25& 26& 34 & 35& 36& 45& 46& 56 \\ \hline
	& & & & & & & & & & & & & & & &\\ \hline 
	${\tilde{\rho}}:=$ &  1 & $e_1$ & 1 & $e_4$ & 0& $e_2$& 0& $e_3$& 1 & $e_2$ &1 &$e_1$ & $e_3$ &1 &$e_4$ \\ \hline
    $s:=$ &  1 &  & 1 &  & 0& & 0& & 1 &  &1 & &  &1 &  \\ \hline
     $s_1:=$ &  1 &$n_1$  & 1 &$n_2$  & 0&$n_1$ & 0&$n_2$ & 1 &  &1 & &  &1 &  \\ \hline
      $s_2:=$ &  1 &$n_1$  & 1 &$n_2$  & 0&$n_1$ & 0&$n_2$ & 1 &{\bf 0}  &1 & &  &1 &  \\ \hline
    $s_3:=$ &  1 &$\Ra 0$  & 1 &$n$  & 0&$\bf1$ & 0&$n$ & 1 &0  &1 &$\Ra 1$ &  &1 &  \\ \hline
    \end{tabular}

     {\sl Table 11: Fact (17) in the context of $m=6$ and $\ell=k=3$}

    We start with a row $s$ that simply copies the 0's and 1's of $\tilde{\rho}$. Due to its 0-bits $s$ avoids the 3-cliques 126,136,146,156,124,234,245,246. The avoidance of 123 and 125 is achieved by $s_1$. Upon\footnote{This is not how the standard $n$-algorithm (glimpsed in 4.3)  proceeds. However, whatever creative way might procure the desired type of $\tilde{\sigma}$ is allowed.} placing {\bf 0} on position 34 one obtains a 012n-row $s_2$ which avoids 134,345,346 all at once. This nudges us to trim $s_2$ to $s_3$ as follows. First, $23\in ones(s_3)$ makes sure that $pos(e_2,e_2)$ gets not swallowed by $zeros(s_3)$, nor by $zeros(s_i)$ for any potential successors $s_i$ of $s_3$. This has a ripple effect:
    $$23\in ones(s_3)\stackrel{n_1}{\Ra}13\in zeros(s_3)\stackrel{e_1}{\Ra} 36\in ones(s_3)$$
    Unfortunately all these efforts were in vain: $s_3$ cannot extend to any 012n-row $\tilde{\sigma}$ that avoids all 3-cliques because $\{23,26,36\}\s ones(s_3)$.

    Each reader modestly familiar with Ramsey numbers could have predicted this: In whatever sophisticated way one chooses the starter row $\tilde{\rho}$ and tries to adapt $\tilde{\sigma}$ to it, one will never achieve
$\tilde{\rho}\cap \tilde{\sigma}\neq\es$ because of (17) and the well-known fact that $R(3,3)=6$.

{\bf 6.5}  What about using (10') for proving ${\cal J}(3,6)\cap{\cal F}(3,6)=\es$? As seen above, the generators
of ${\cal F}(3,6)$ are the minimal hitting sets $G_1,...,G_s$ of the hypergraph ${\HH}_{3,6}$. Dually the facets $F_1,...,F_t$ of ${\cal J}(3,6)$ are the maximal noncovers of the hypergraph ${\HH}_{3,6}$. It turns out that $s=t=211$ and hence the naive application of (10') demands to ckeck $211^2$ pairs $G_i,F_j$.

\begin{center}
   \includegraphics[scale=0.8]{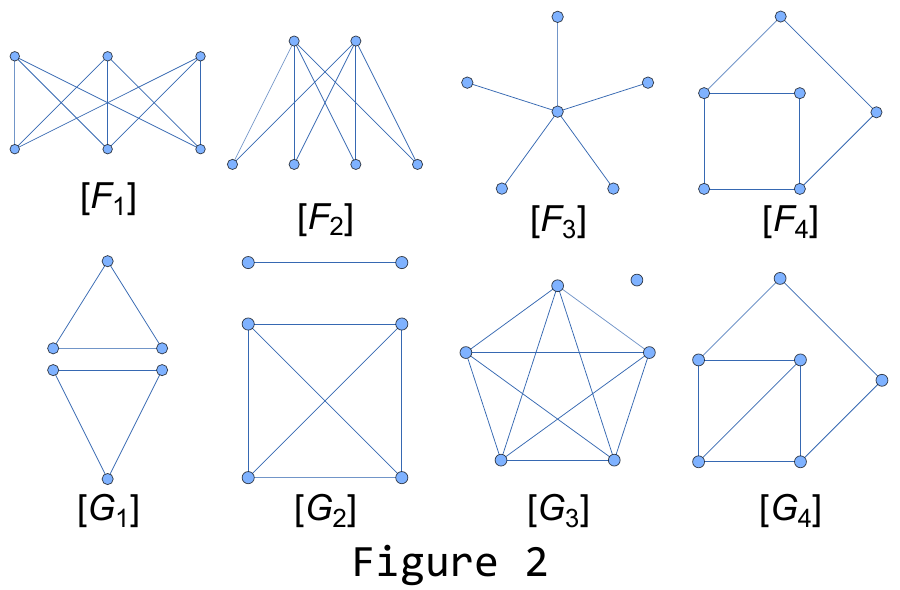} 
\end{center}

Fortunately one can do better. Namely, there are just four isomorphy classes of facets $F_j$, w.l.o.g. we can assume they are $[F_1],[F_2],[F_3],[F_4]$. Their respective cardinalities are 10,15,6,180 (which sum up to 211). Likewise the generators $G_i$ come in four isomorphy classes $[G_1],[G_2],[G_3],[G_4]$; see Figure 2. 
As opposed to checking $211^2$ pairs, the task has simplified significantly: Check that none of the four (unlabeled) graphs $[G_i]$ extends to any of the four graphs $[F_j]$ ("extends" refers to the respective edge-sets).
This is immediate because (by construction) each $[G_i]$ lacks 3-anticliques, while (by\footnote{Generally a facet of ${\cal F}_{\ell,n}$ has no $\ell$-clique, but it may or may not have $k$-anticliques (for various $k$).} coincidence) all $F_j$ have 3-anticliques. Of course one cannot go from "without 3-anticliques" towards "with 3-anticliques" by adding edges.

Among the three real-life (or: real-mathematics) applications presented in Sections 5,6,7, it is Section 6 that needs the most extra input. Specifically, expertise concerning symmetry exploitation and/or random sampling  will be required to refine the sketched ideas.

\section{Fixpoints of Monotone Boolean Networks  }

A digraph $D$ with (finite) vertex set $S$ is called {\it functional} if each vertex has out-degree 1. Put $\Phi(a):=b$ if there is an arc from vertex $a$ to vertex $b$. This yields a well-defined selfmap $\Phi:S\to S$. Conversely each selfmap derives from a unique functional digraph $D$. As is well known, each connected component of $D$ (viewed as undirected graph) consists of a directed cycle  which possibly\footnote{Planted trees never occur iff $\Phi$ is bijective; this is the familiar cycle representation of permutations.} has directed trees "planted" upon it. The 1-cycles (i.e. loops)  match the fixpoints $a=\Phi(a)$ of our selfmap. Some of the most applicable selfmaps have the form $\Phi:\{0,1\}^n\to\{0,1\}^n$ and are called {\it Boolean networks}. They were introduced 1969 by Stuart Kauffman who propagated them as a simple model for gene regulatory networks.  Today Google Scholar finds 1.2 million results for the key words "Boolean network". Some of the established jargon in this field: attractor := directed cycle, singleton attractor := fixpoint, attractor basin := connected component, Garden of Eden := vertex without predecessors, i.e. of indegree 0.

One major theme is the identification of singleton attractors. 
While it is NP-complete to decide the existence of  fixpoints, plenty quite different methods (surveyed in [MA]) have been propagated to detect them.  Finding a Boolean formula whose models are exactly the fixpoints of $\Phi$ is just one of these methods, but it is the one we expand upon in  Subsections 7.2 to 7.4. (Subsection 7.1 concerns preliminaries that we could have stated in Section 4 already.)

 {\bf 7.1} Recall (Table 5) that the modelset of a pure {\tt Horn-CNF} (i.e. $CS(\Sigma)$ for some implication family $\Sigma$) can be rendered as disjoint union of $012n$-rows $\sigma_j$. The same holds when "pure  {\tt Horn-CNF}" is generalized to "{\tt Horn-CNF}"; in [W2] this generalization (which essentially blends  the implication $n$-algorithm with the noncover $n$-algorithm) is called the Horn $n$-algorithm. Dually the AntiHorn $e$-algorithm renders the modelset of each  {\tt AntiHorn-CNF} as disjoint union of $012e$-rows $\rho_i$. It follows that the modelset of each CNF of type {\tt Horn$\wedge$AntiHorn} can be calculated with the $\rho,\sigma$-{\it mechanism}, i.e. by repeated enumeration of set-families $\rho_i\cap\sigma_j$.(This will be illustrated in detail in Table 12.)
 Theorem 2 below concerns proper  {\tt Horn$\wedge$AntiHorn}, thus not the subtype  {\tt Positive$\wedge$Negative} which was prominent in Sections 5 and 6.

{\bf 7.2}  Formally a {\it Boolean network} is a map $\Phi:\{0,1\}^m\to\{0,1\}^m,\ x\mapsto(\Phi_1(x),...,\Phi_m(x))$, such that each {\it component function} $\Phi_i:\{0,1\}^m\to\{0,1\}$ is a  Boolean function. Observe that $y\in\{0,1\}^m$ is a  fixpoint (i.e. $\Phi(y)=y$) iff $y_i=\Phi_i(y)$ for all $1\le i\le m$. Let us construct some  CNF $f$ such that $Mod(f)$ coincides with the set of all fixpoints of $\Phi$.

 To fix ideas, we focus on $i:=3$ and suppose $\Phi_3(x):=x_1\wedge(x_2\vee x_4)\wedge(x_4\vee x_5)$. Coupled to $\Phi_3$ consider this CNF of type {\tt Horn$\hspace{2pt}\wedge\hspace{2pt}$AntiHorn}:
$$f_3:=\Big[(\ol{y_3}\vee x_1)\wedge(\ol{y_3}\vee x_2\vee x_4)\wedge(\ol{y_3}\vee x_4\vee x_5)\Big]\
\wedge\ \Big[(y_3\vee\ol{x_1}\vee\ol{x}_4)\wedge (y_3\vee\ol{x_1}\vee\ol{x}_2\vee\ol{x}_5)\Big]$$
We claim that for all $y:=(x_1,x_2,x_4,x_5,y_3)\in\{0,1\}^5$ it holds that
$$(18)\quad y_3=\Phi_3(x)\ \LRa\ y\in Mod(f_3)$$
{\it Proof of (18).} As to "$\Ra$",  suppose first that $y_3=0$. Then the first three clauses of $f_3$ are satisfied because $\ol{y}_3=1$. The other two clauses of $f_3$ are satisfied because by assumption $0=\Phi_3(x)=
x_1\wedge(x_2\vee x_4)\wedge(x_4\vee x_5)$, and so $x_1=0$ or $x_2=x_4=0$ or $x_4=x_5=0$. Suppose now that $y_3=1$.
Then the last two clauses of $f_3$ are satisfied. By assumption $1=\Phi_3(x)=x_1\wedge(x_2\vee x_4)\wedge(x_4\vee x_5)$, and so $x_1=1$ and ($x_2=1$ or $x_4=1$) and ($x_4=1$ or $x_5=1$). This guarantees that the first three clauses of $f_3$ are satisfied as well.

As to  "$\Leftarrow$" in (18), we assume that $y\in Mod(f_3)$ and make again a case distinction. {\bf Case 1:} $y_3=0$. Since $f_3(y)=1$ by assumption, it holds in particular that $g:=(\ol{x}_1\vee\ol{x}_4)\wedge (\ol{x}_1\vee\ol{x}_2\vee\ol{x}_5)=1$. An easy calculation (more background later) shows that 
$g=\ol{x}_1\vee (\ol{x}_2\wedge\ol{x}_4)\vee (\ol{x}_4\wedge\ol{x}_5)$.
Therefore either $\ol{x}_1=1$ or $\ol{x}_4=\ol{x}_2=1$ or $\ol{x}_4=\ol{x}_5=1$.
It is evident that
$\ol{x}_1=1\Ra x_1=0\Ra \Phi_3(x)=0\ (=y_3)$, that $\ol{x}_2=\ol{x}_4=1\Ra x_2=x_4=0\Ra\Phi_3(x)=0$, and that
$\ol{x}_4=\ol{x}_5=1\Ra x_4=x_5=0\Ra\Phi_3(x)=0$. {\bf Case 2:} $y_3=1$. From $f_3(y)=1$ follows in particular that $g':=x_1\wedge(x_2\vee x_4)\wedge(x_4\vee x_5)=1$ One checks that $g'=(x_1\wedge x_4)\vee(x_1\wedge x_2\wedge x_5)$. Therefore $x_1=x_4=1$ or $x_1=x_2=x_5=1$. Clearly $x_1=x_4=1\Ra \Phi_3(x)=1\ (=y_3)$, and likewise
 $x_1=x_2=x_5=1\Ra \Phi_3(x)=1$. This proves (18).

 {\bf 7.3}  In this  Subsection we focus on {\it monotone}\footnote{These arise frequently in biology.} Boolean networks in the sense that all component functions must be positive Boolean functions. 

 {\bf Theorem 2: }{\it For each monotone Boolean network $\Phi$ there exists a CNF $f$ of type {\tt Horn$\wedge$AntiHorn} such that the
 fixpoints of $\Phi$ are exactly the models of $f$. Furthermore $f$ has the same  variables but more clauses than $\Phi$.}
 
{\it Proof.} A moment's thought shows that the positivity of  $\Phi_3$ above was crucial\footnote{Actually, by duality one could replace "positive" by "negative". This also shows that in the statement of Theorem 2 one can  interpret "monotone" in the broader sense that each component function must either be positive or negative.} for $f_3$ being of type {\tt Horn$\wedge$AntiHorn}. Generally for all $1\le i\le m$ one can set up a CNF $f_i$ of type {\tt Horn$\wedge$AntiHorn} such that the analogon of (18) holds. Therefore, if we define $f:=f_1\wedge f_2\wedge...\wedge f_m$, then $f$ stays of type {\tt Horn$\wedge$AntiHorn} and is such that for all $y\in\{0,1\}^m$ it holds that
$y=\Phi(y)\ \LRa\ y\in Mod(f).$ $\square$

The part of Theorem 2  stating that "$f$ has more clauses than $\Phi$" can potentially be refined. Namely, observe that while the first part $\big[....\big]$ in the definition of $f_3$ clearly mimicks $\Phi_3(x)$, the second part $\big[....\big]$ seems enigmatic, but let us see how it comes about. It is the CNF triggered by some DNF (i.e. $\ol{x}_1\vee (\ol{x}_2\wedge\ol{x}_4)\vee (\ol{x}_4\wedge\ol{x}_5)$) which matches $\Phi_3(x)$ in obvious ways. If we had an upper bound for the lengths of all these triggered CNF's ($1\le i\le m$), then "$f$ has more clauses than $\Phi$" could be made concrete.

  {\bf 7.4} According to [MA] the {\it indegree} $K$ of a (not necessarily monotone) Boolean network $\Phi$ is the maximum number of (essential) variables that a component function can have. Thus if $K=2$ and all component functions are in CNF format, then each component function is of type $x_i=\ell_j$ or $x_i=\ell_j\vee \ell_k\  (possibly\ i\in\{j,k\})$. Here $\ell_j$ is a {\it literal}, i.e either $x_j$ or $\ol{x}_j$, and likewise for $\ell_k$. Leaving the easier case
$x_i=\ell_j$ aside, we claim that
$$(19)\quad x_i=\ell_j\vee\ell_k\  \LRa\ (\ol{x}_i\vee \ell_j\vee\ell_k)\wedge(x_i\vee\ol{\ell}_j)\wedge (x_i\vee\ol{\ell}_k)=1$$
The proof is similar to (parts of) the proof of (18). The crucial fact is that the right hand side in (19) is a 3-CNF, i.e. a CNF all of whose clauses have length at most 3. As is pointed out in [MA], it follows that for each Boolean network $\Phi:\{0,1\}^m\to\{0,1\}^m$ of indegree 2 there is a 3-CNF $f$ with at most $3m$ clauses such that the fixpoints of $\Phi$ are the models of $f$. In particular, the existence of a $\Phi$-fixpoint can be decided in time $O(1.3303^m)$ due to a result of [MTY] about 3-CNFs. Unfortunately this pleasant state of affairs does not carry over to Boolean networks with indegree $K>2$ since they trigger (according to [MA]) a CNF $f$ with $2^{K+1}m$ clauses and furthermore instead of $3-SAT$ one has to solve $(K+1)-SAT$.

{\bf 7.5} We mention in passing that idempotent monotone Boolean networks $\Phi$ (so $\Phi\circ\Phi=\Phi$) have been characterized in [RW,p.150]. Hence these  $\Phi$'s have plenty fixpoints, each image $\Phi(x)$ is one! The author does not know whether this property is relevant in the biological sciences, but it is so in the field of nonlinear image filtering. That is\footnote{Previously we wrote $\{0,1\}^m$ instead of $\{0,1\}^{[m]}$, where $[m]:=\{1,2,...,m\}$.} because $\{0,1\}^{[m]}$  can be replaced by $\{0,1,2,..,t\}^{[m]\times [n]}$, and each member of  $\{0,1,2,..,t\}^{[m]\times [n]}$ can be viewed as image of dimension $m\times n$ with $mn$ pixels, having greytones varying from $0$ (=white) to $t$ (=black).

\section{The benefit of turning arbitrary CNF's to type {\tt Horn$\wedge$AntiHorn}}

We stick to {\tt Horn$\hspace{2pt}\wedge\hspace{2pt}$AntiHorn}, but in a vein unrelated to Boolean networks. To begin with,
   each CNF $f'$ can be transformed to some {\it equisatisfiable} 3-CNF $g'$ (i.e. $f'$ is satisfiable iff $g'$ is).  Therefore 3-CNF's are called {\it universal}\footnote{ Since each 3-CNF is evidently of type
{\tt Horn$\hspace{2pt}\wedge\hspace{2pt}$AntiHorn}, also {\tt Horn$\hspace{2pt}\wedge\hspace{2pt}$AntiHorn} is universal.}.  Unfortunately, because transforming  CNF's to 3-CNF's  inflates the size of the latter, this manipulation is "only" of theoretic interest (but very much so, see 8.1). 
  In Theorem 3 of 8.2 we prove a less inflationary
transformation from an arbitrary CNF $f'$  to some {\tt Horn$\hspace{2pt}\wedge\hspace{2pt}$AntiHorn} $g'$. Once
$Mod(g')$ has been  calculated  (e.g. with the $\rho,\sigma$-mechanism, as illustrated in 8.3), the modelset of interest, i.e.
$Mod(f')$,  can  be retrieved from $Mod(g')$ in natural ways (as illustrated in 8.4).

\vspace{5mm}

{\bf 8.1} As mentioned, with each CNF $f$ one can associate a 3-CNF $f'$ that is equisatisfiable. In brief  [GJ,p.48], each clause of $f$ of length $k>3$ triggers $k-2$ many new 3-clauses of $f'$ (which together feature $k-3$ new variables, i.e. not occuring in $f$). Thus $f'$ has many more  variables and clauses than $f$.  The sketched transformation shows that the NP-completeness of satisfiability (of arbitrary CNFs) carries over to the NP-completeness of 3-satisfiability. Why is this important? Because 3-CNFs are crisper than general CNFs, they are better suited to establish further NP-completeness results. One example is deciding the presence of fixpoints in Boolean networks (Section 7). As another example, in [GJ,p53] it is shown how each instance of 3-satisfiability can be reduced to an instance of "vertex cover", proving that "vertex cover" is NP-complete.
In turn "vertex cover" can then be used to show the NP-completeness of many other combinatorial problems. While 3-CNFs are essential theoretic tools (also in 7.4), the author is not aware\footnote{Information to the contrary is welcome. On the other hand, as is well known, 2-CNFs  have plenty real world applications.}  of {\it real world} applications that systematically (not just incidentally) lead to  3-CNFs.

{\bf 8.2} Let us embark on the less inflationary reduction of general CNFs $f$ to type {\tt Horn$\wedge$AntiHorn}  CNFs. It is modelled on an idea which is sketched  (yet not formally verified) in [G] and which reduces  arbitrary CNFs $f$ to type {\tt Positive$\wedge$Negative}. In [G] {\it each} clause $C$ of $f$ that fails to be {\tt positive} or {\tt negative} triggers (i) a  new variable, and (ii) replaces $C$ by two suitable clauses $C',C''$. Our approach does the same, yet only for the {\it fewer} clauses $C$ of $f$ that fail to be {\tt  Horn} or {\tt AntiHorn}.

{\bf 8.2.1} Let us carry out the details on a sufficiently rich toy example. Thus consider 
$$(20)\ F=F_0\wedge F_1\wedge F_2:=\Big[(x_1\vee x_2\vee x_3\vee\ol{x}_4)\wedge (\ol{x}_1\vee \ol{x}_3\vee x_5\vee\ol{x}_6)\Big]\wedge \Big[\ol{x}_2\vee x_4\vee\ol{x}_5\vee x_6\Big]\wedge \Big[x_1\vee \ol{x}_2\vee \ol{x}_4\vee x_5\vee\ol{x}_6\Big]$$
Generally $F_0$ comprises the clauses that are either {\tt  Horn} or {\tt AntiHorn}, and $F_1,...,F_s$ are the clauses which are neither (here $s=2$). This triggers a certain {\tt Horn$\wedge$AntiHorn} CNF $G=G_0\wedge G_1\wedge...\wedge G_s$ as follows. First, $G_0:=F_0$ is the part of $F$ which is {\it already} {\tt Horn$\wedge$AntiHorn}.
Furthermore each 'bad' clause $F_i\ (i\ge 1)$ triggers one new variable, say $x_k$, and two clauses $G_i^0$ and $G_i^1$. Namely, the literals of $G_i^0$ are $\ol{x}_k$ together with the positive literals in $F_i$, and
the literals of $G_i^1$ are $x_k$ together with the negative literals in $F_i$. The above-mentioned parts $G_i$ are defined as $G_i:=G_i^0\wedge G_i^1\ (1\le i\le s)$. For $F$ in (20) we therefore get $G$ in (21):
$$(21)\quad G=G_0\wedge G_1\wedge G_2:=\Big[(x_1\vee x_2\vee x_3\vee\ol{x}_4)\wedge (\ol{x}_1\vee \ol{x}_3\vee x_5\vee\ol{x}_6)\Big]$$
$$\wedge\ \Big[(x_4\vee x_6\vee\ol{x}_7)\wedge (\ol{x}_2\vee\ol{x}_5\vee x_7)\Big]\wedge \Big[(x_1\vee x_5\vee \ol{x}_8)\wedge (\ol{x}_2\vee \ol{x}_4\vee \ol{x}_6\vee x_8)\Big]$$

{\bf Theorem 3: }{\it Let $F':\{0,1\}^m\to\{0,1\}$ be any Boolean function in CNF-format, and let $s$ be the number  of clauses which are neither {\tt Horn} nor {\tt AntiHorn}. Let $G':\{0,1\}^{m+s}\to\{0,1\}$ be defined by the {\tt Horn$\wedge$AntiHorn} CNF that is obtained from the CNF of $F'$ in a fashion analogous to the way (21) was derived from (20). It then holds that
$Mod(F')$ equals\\
$U:=\big\{(y_1,..,y_m)\in\{0,1\}^m:\ (\exists y_{m+1},..,y_{m+s}\in\{0,1\})\ (y_1,..,y_m,y_{m+1},..,y_{m+s})\in Mod(G')\big\}$.}

In brief, Theorem 3 says: The auxiliary function $G'$ has $m+s$ variables and $Mod(F')$ is the {\it projection} of $Mod(G')$ onto the first $m$ variables. In particular $Mod(F')$ is empty iff $Mod(G')$ is empty (see also 8.5).

{\it Proof.} For purposes of exposition we take $m:=6, s=2$ and $F':=F,\ G':=G$ from (20),(21) above.
In order to prove the inclusion $\s$ in the claim $Mod(F)=U$ take any $(y_1,..,y_6)\in Mod(F)$. From $\ol{y}_2\vee y_4\vee \ol{y}_5\vee y_6=1$ follows that $y_4\vee  y_6=1$ or $\ol{y}_2\vee \ol{y}_5=1$ (or both). Hence either
$(y_4\vee y_6\vee\ol{1})\wedge (\ol{y}_2\vee\ol{y}_5\vee {\bf 1})=1$ or $(y_4\vee y_6\vee\ol{0})\wedge (\ol{y}_2\vee\ol{y}_5\vee {\bf 0})=1$.  If the former takes place then $(y_1,..,y_6,y_7):=(y_1,..,y_6,{\bf 1})\in Mod(G_1)$. If the latter takes place then $(y_1,..,y_6,y_7):=(y_1,..,y_6,{\bf 0})\in Mod(G_1)$. Since $(y_1,..,y_6)\in Mod(F)\s Mod(G_0)$ by assumption, this implies that
$$(22)\quad  (y_1,..,y_6,y_7):=(y_1,..,y_6,0)\in Mod(G_0\wedge G_1)\ or\ (y_1,..,y_6,y_7):=(y_1,..,y_6,1)\in Mod(G_0\wedge G_1)$$
Likewise it follows from $y_1\vee\ol{y}_2\vee\ol{y}_4\vee y_5\vee\ol{y}_6=1$ that either
$(y_1,..,y_6,y_8):=(y_1,..,y_6,0)\in Mod(G_0\wedge G_2)$ or $(y_1,..,y_6,y_8):=(y_1,..,y_6,1)\in Mod(G_0\wedge G_2)$.
In view of (22) we conclude that at least one of the bitsrings 
$(y_1,..,y_6,0,0),(y_1,..,y_6,0,1),(y_1,..,y_6,1,0),(y_1,..,y_6,1,1)$ belongs to $Mod(G_0\wedge G_1\wedge G_2)=Mod(G)$. This amounts to say that $(y_1,...,y_6)$ belongs to $U$.

In order to prove the inclusion $\supseteq$ in the claim $Mod(F)=U$,
take $(y_1,..,y_6)\in U$. By definition of $U$ there are $y_7,y_8\in\{0,1\}$ with 
$(y_1,..,y_6,y_7,y_8)\in Mod(G)$. (Then necessarily $(y_1,..,y_6,y_7)\in Mod(G_0\wedge G_1)$ and
$(y_1,..,y_6)\in Mod(G_0)$.)
We need to show that $(y_1,..,y_6)\in Mod(F)$. So far we only know that $(y_1,..,y_6)\in Mod(F_0)$ (because $F_0=G_0$). To fix ideas assume that $(y_1,..,y_6,y_7,y_8)=(y_1,..,y_6,0,1)$ (the cases with 01 replaced by 00,10,11 respectively are analogous). From $(y_1,..,y_6,y_7)\in Mod(G_1)$ follows $(y_4\vee y_6\vee\ol{0})\wedge (\ol{y}_2\vee\ol{y}_5\vee 0)=1$, hence $\ol{y}_2\vee\ol{y}_5=1$, hence $\ol{y}_2\vee y_4\vee\ol{y}_5\vee y_6=1$,
hence $(y_1,..,y_6)\in Mod(F_1)$. Similarly from $(y_1,..,y_6,y_7,y_8)\in Mod(G_2)$ follows
that $(y_1\vee y_5\vee\ol{1})\wedge (\ol{y}_2\vee \ol{y}_4\vee \ol{y}_6\vee 1)=1$, hence $y_1\vee y_5=1$, hence
$y_1\vee \ol{y}_2\vee \ol{y}_4\vee y_5\vee \ol{y}_6=1$, hence $(y_1,..,y_6)\in Mod(F_2)$. Since the latter bitstring also lies in $Mod(F_0)$ and $Mod(F_1)$, one concludes $(y_1,..,y_6)\in Mod(F)$.
 $\square$

{\bf 8.3} Here we carry out the $\rho,\sigma$-mechanism introduced in 7.1 by explicitely representing $Mod(G)$ as disjoint union of parts $\rho_i\cap\sigma_j$,
where the 012e-rows $\rho_i$ derive from the {\tt AntiHorn} part of $G$, i.e. from
$$(23)\quad (x_1\vee x_2\vee x_3\vee\ol{x}_4)
\wedge\ (x_4\vee x_6\vee\ol{x}_7)\wedge (x_1\vee x_5\vee \ol{x}_8),$$
and the 012n-rows $\sigma_j$ from the {\tt Horn} part of $G$, i.e. from
$$ (24)\quad (\ol{x}_1\vee \ol{x}_3\vee x_5\vee\ol{x}_6)
\wedge (\ol{x}_2\vee\ol{x}_5\vee x_7)\wedge (\ol{x}_2\vee \ol{x}_4\vee \ol{x}_6\vee x_8).$$

{\bf 8.3.1} Specifically, like Table 5 also Table 12 relies on a LIFO stack. Its first two members are the top row $(e,e,e,1,2,2,2,2)$ and the bottom row $(2,2,2,0,2,2,2,2) $ (blanks amount to don't-care 2's). By construction all bitstrings in the top row are models of the first clause in (23) and, incidentally, of the second clause. Therefore the 3rd clause is pending (to be imposed). In the bottom row the 2nd clause of (23) is pending. After a few more steps one obtains the 012e-rows $\rho_1$ to $\rho_7$ whose disjoint union is the modelset of formula (23).
Likewise the implication $n$-algorithm calculates the 012n-rows $\sigma_1,...,\sigma_6$ whose disjoint union is the modelset of the {\it Horn} formula (24).

{\bf 8.3.2} Section 9 will be  devoted to the systematic computation of intersections of 012e-rows with 012n-rows. But
in our toy example the 42 intersections $\rho_i\cap\sigma_j$ can be determined ad hoc; in particular 21 of them happen to be empty because of 0-1 clashes. How to get, say,  $\rho_5 \cap\sigma_1$ ad hoc? Because of $5\in zeros(\sigma_1)$ the second $e$ in $\rho_5$ is turned to $0$, which forces the first $e$ to $1$. It follows that 
$\rho_5 \cap\sigma_1=({\bf 1},2,2,0,{\bf 0},1,2,1)\cap\sigma_1=:\rho_5' \cap\sigma_1$. Since $ones(\rho_5')$ covers the first and third $n$ in $\sigma_1$, the middle $n$ turns to $0$ and one concludes that $\rho_5 \cap\sigma_1=(1,2,{\bf 0},0,0,1,2,1)$. 

\begin{tabular}{l|c|c|c|c|c|c|c|c|c|c|c|c|c|c|c|c|c||c|c|}
&1 &2 &3 &4 &5 &6 &7 &8 & & & &1 &2 &3 &4 &5 &6 &7 &8   \\ \hline
	& & & & & & & & & & & & & & &  &  & & &\\ \hline
    &$e$ & $e$&$e$ &  \bf 1 & &  &  &  & pend. 3 & &  &  & &  & &  &  &  &   \\ 
     &  & &  &  \bf 0 & &  &  &  & pend. 2 & &  &  & &  & &  &  &  &   \\ \hline
	$\rho_1=$ &  \bf 1 & &  &  1 & &  &  &  & final & & $\rho_1\cap\sigma_1=$ & 1 & &  $n$ &1 &0  & $n$ &  &1 \\ 
	$\rho_2=$ &  \bf 0 & $e$& $e$ &  1 &\bf 1 &  &  &  & final & & $\rho_1\cap\sigma_2=$ & 1 & & &1 &0&0 &  &0 \\
    $\rho_3=$ &  \bf 0 &$e$ &$e$  &  1 &\bf 0 &  &  & 0 & final & & $\rho_1\cap\sigma_3=$ & 1 &0 & 0  &1 &0&1 & &0 \\
     &   & &  &  0 & &  &  &  & pend. 2 & & $\rho_1\cap\sigma_4=$ & 1 &0 &  &1 &1 & &  & \\ \hline
      &  & &  &  0 & & \bf 1 &  &  & pend. 3 & & $\rho_1\cap\sigma_5=$ & 1 &1 &  &1 &1  & 0 & 1 &0 \\
       &  & &  &  0 & & \bf 0 & 0 &  & pend. 3 & & $\rho_1\cap\sigma_6=$ & 1 & 1&  &1 &1  &  &1  &1 \\ \hline
   $\rho_4=$  &  & &  &  0 & &  1 &  & \bf 0 & final & & $\rho_2\cap\sigma_4=$ & 0 & 0& 1 &1 &1  &  &  & \\
   $\rho_5=$  & $e$ & &  & 0 &$e$&  1 &  & \bf 1 & final & & $\rho_2\cap\sigma_5=$ & 0 &1 &  &1 &1  & 0 &1  &0 \\
   &  & &  &  0 & &  0 & 0 &  & pend. 3 & & $\rho_2\cap\sigma_6=$ &0 &1 & &1 & 1 & & 1 &1 \\ \hline
 $\rho_6=$ &  & &  &  0 & &  0 & 0 &\bf 0  & final & & $\rho_3\cap\sigma_2=$ & 0 &$e$& $e$ &1 &0  &0 &  &0 \\
 $\rho_7=$ &$e$  & &  &  0 &$e$ &  0 & 0 &\bf 1  & final & & $\rho_3\cap\sigma_3=$ &0 &0 & 1 &1 &0  &1 &  &0 \\ \hline
& & & & & & & & & & & $\rho_4\cap\sigma_3=$  &$n$ & &$n$ &0& 0 &1 & &0 \\ \hline

 & $n$ & &$n$ &  &\bf 0 & $n$  &  &  & pend. 3 & &$\rho_4\cap\sigma_4=$ &  &0 &  &0 &1 &1  &  & 0  \\ 
  &  & & &  &\bf 1 & &  &  & pend. 2 & &$\rho_4\cap\sigma_5=$  &  &1 &  &0 &1  &1  &1  &0   \\\hline
$\sigma_1=$ & $n$ & &$n$ &  & 0 & $n$  &  &\bf 1  & final & & $\rho_5\cap\sigma_1=$ &1& &0  &0 &0  &1  &  &1   \\ 
$\sigma_2=$ & & & &  & 0 & \bf 0  &  &\bf 0  & final & &$\rho_5\cap\sigma_4=$& &0 & &0 &1  &1  &  &1   \\ 
$\sigma_3=$ &$n_1$ &$n_2$ &$n_1$ &$n_2$ & 0&\bf 1&  &\bf 0& final & &$\rho_5\cap\sigma_6=$ & &1 & &0 &1 &1 &1 &1  \\ 
  &  & & &  & 1 & &  &  & pend. 2 & &$\rho_6\cap\sigma_2=$  &  & &  &0 &0  &0  &0  &0   \\  \hline
$\sigma_4=$   &  &\bf 0 & &  & 1 & &  &  & final & &$\rho_6\cap\sigma_4=$  &  & 0&  &0 &1  &0  &0  &0   \\
$\sigma_5=$  &  &\bf 1 & &$n$  & 1 &$n$ &1  &\bf 0  & final & &$\rho_7\cap\sigma_1=$  & 1 & & &0 &0 &0 &0  &1   \\
$\sigma_6=$  &  &\bf 1 & & & 1 & &1  &\bf 1 & final & &$\rho_7\cap\sigma_4=$  &  &0 &  & 0&1 &0&0  &1   \\ \hline
 \end{tabular}
 
 {\sl Table 12: Calculating all models of {\tt Horn$\hspace{2pt}\wedge\hspace{2pt}$AntiHorn} formulas}

{\bf 8.4} According to Theorem 3 the modelset of $F$ in (20) is the union of the rows $\theta_1,...,\theta_{21}$ obtained by cutting the last $s=2$ components of the 21 rows $\rho_i\cap\sigma_j$ listed in Table 12. A slight drawback is the fact that this union needs not be disjoint. For instance $\theta_3\s\theta_1$.
In the general case the last $s$ components of $\rho_i\cap\sigma_j$ may either feature $e$ symbols or\footnote{But not both, as will be explained in Section 9.} $n$-symbols. Let thus $\stackrel{\ra}{e}$ have length $k$ and let $k_0\in \{1,2,..,k\}$ be the number of $e$-symbols among the last $s$ components of $\rho_i\cap\sigma_j$. A moment's reflection shows that upon cutting these  $s$ components the $k-k_0$
surviving $e$'s of $\stackrel{\ra}{e}$ simply turn to $2$'s. (Likewise for cut wildcards $\stackrel{\ra}{n}$.) What about the fact that the cut rows, call them $\theta_i$, may not be disjoint? This may not be disturbing\footnote{Even the non-disjoint format suffices for random sampling, and the sum of all cardinalities $|\theta_i|$  is an upper bound for $|Mod(F)|$.}. If however disjointness is desired, then first replace  each  012e-row $\theta_i$ by a disjoint union of 012-rows (as shown in 9.7.1). Of course a 012-row originating from $\theta_i$ need not be disjoint from a 012-row originating from $\theta_j$. In order to achieve overall disjointness proceed as in Section 2.

{\bf 8.5} If in Theorem 3 one is only interested to decide the satisfiability of $F'$ (i.e. $Mod(F')\stackrel{?}{=}\es$) then the issue of cut rows evaporates altogether. Rather we are brought back to the {\it old asymmetry at the heart of automated reasoning }(Section 1). Specifically, the satisfiability of $F'$ can be reduced to the satisfiability of $G'$ which, it being of type {\tt Horn$\wedge$AntiHorn}, amounts to the interplay of 012e-rows $\rho_i$ with 012n-rows $\sigma_j$ (according to Section 4 they can be generated fast).  Recall, the satisfiability of $G'$ (and thus of $F'$) amounts to the existence of {\it some} pair $(i,j)$ with $\rho_i\cap\sigma_j\neq\es$. Unsatisfiability amounts to certify that $\rho_i\cap\sigma_j=\es$ for {\it all} pairs $(i,j)$. While there may be plenty pairs $(i,j)$ (e.g. in Section 6), and so the certificate may be long,
the certificate is crisp\footnote{That is, for each individual pair $(i,j)$ showing that $\rho_i\cap\sigma_j=\es$ boils down to identifying a trivial reason (3.3) either immediately or after a brief 0,1,2-propagation.} and easily subdivided in digestible (and distributable) pieces. 

This prompts us to have a closer look at [HK] which was previously mentioned in the context of a 200 terabyte certificate of unsatisfiability. Article [HK] surveys the history of unsatisfiability proofs in the last 25 years. The emphasis is on clausal proofs, with  a nice toy example being provided in [HK,Fig.2]. What concerns the CNF  $F'$ that triggered the  200 terabyte certificate (=clausal proof), $F'$ actually is of type {\tt Positive$\wedge$Negative} already, thus $G'=F'$ in the terminology of Theorem 3.   Furthermore, the negative part is dual to the positive part. Therefore, if one were to expand the positive part into $R$ many 012e-rows $\rho_i$, upon switching 0's and 1's, as well as $e$-symbols and $n$-symbols, one gets the $R$ many 012n-rows $\sigma_j$ for free.  While the number $R^2$ of pairs $(\rho_i,\sigma_j)$ might be intimidating\footnote{Applying a  "look ahead strategy", akin to the kind described in [HK], yields a decent order in which the clauses are fed to the $e$-algorithm in Section 4. This trims $R$ and hence $R^2$.}, storing these pairs almost surely takes way less than 200 terabytes of space. Whether establishing unsatisfiability by pointing out $R^2$ trivial reasons takes less than 4 CPU years, is another issue. Whatever time it takes, as testified in Table 2', the {\it storing} of the $R^2$ trivial reasons takes at most $m$ times the space of storing the (untreated) $R^2$ pairs $(\rho_i,\sigma_i)$. We speculate that such $\rho,\sigma$-proofs of unsatisfiability sometimes take less space than clausal proofs. 

Last not least, [HK] also expands (like us in Section 6) about Ramsey numbers.

\section{The missing proof}

The purpose of 9.1 and 9.2 is to convince the reader that  a simple method to find a bitstring
in a set of type $(2e-row)\cap (2n-row)$ probably does not exist.
This justifies the techniques in 9.3 and 9.4 that ultimately establish the pending proof of Theorem 1 which was stated in Subsection 3.6. Based on Theorem 1 we show (in Theorem 5 of 9.5) that {\it all} bitstrings in a set of type $(2e-row)\cap (2n-row)$ can be enumerated in output-polynomial time. Subsection 9.6 is dedicated to the variant $(2g-row)\cap (2n-row)$, and in 9.7 we give an alternative proof of Theorem 5 which is based on the two questions left open in 3.8.

{\bf 9.1}  Refering to Table 13,
we  step-wise build a bitstring $y=(y_1,...,y_{24})\in\tau_0\cap\sigma_0$, starting by setting any component $y_i$ to 1. For instance, let the {\it seed} be $y_1:=1$. Because $y_1$ happens to lie in  $\stackrel{\ra}{e_1}$, this $e$-wildcard with
$pos(\stackrel{\ra}{e_1})=\{1,2,3,4\}$ is now {\it calmed} (Step 1)
in the sense that we can  set $y_2,y_3,y_4$ freely (indicated by smileys). Putting
$y_2=y_3=y_4:=0$ is a good choice because in turn it  calms $\stackrel{\rightarrow}{n_1},\stackrel{\rightarrow}{n_7},\stackrel{\rightarrow}{n_9}$ (Step 2). Thus {\it each} future bitstring
of type $y=(1,0,0,0,...)$ will satisfy these three $n$-wildcards of $\sigma_0$. Furthermore, e.g. the fact that $\sigma_0[5]=\sigma_0[10]=n_9$ make 5 and 10 {\it inviting positions} in the sense that  $y_5,y_{10}$ can be chosen risk-free as 0 or 1 in the future. Several other inviting positions are indicated by smileys as well (Step 2). At this stage five pending $e$-wildcards $\stackrel{\ra}{e}$ carry smileys. It makes sense to set exactly one smiley per wildcard $\stackrel{\ra}{e}$ to $1$ and the other components to $0$. This simultaneously calms $\stackrel{\ra}{e}$ and the 0's may create new inviting positions; in fact two new ones are created (Step 3).
In the same way as Step 2 triggered Step 3, now Step 3 triggers Step 4, and Step 4 triggers Step 5.
At this stage no more inviting positions are on offer, and so the procedure halts. We pinned down all $m=24$ positions, i.e. found a bitstring $y\in \tau_0\cap\sigma_0$ (spelled out in Table 13). However, observe that planting the seed $y_1:=1$ (or in fact any seed $y_j:=1$) was a risky operation, because it is not guaranteed that the wildcard $\stackrel{\ra}{n_2}$ will ever be calmed; we were just lucky that $\stackrel{\ra}{n_2}$ incidentally got calmed in Step 4.

\begin{tabular}{l|c|c|c|c|c|c|c|c|c|l}
	& $1\ \ 2\ \ 3\ \ 4$ & $5\ \ 6\ \ 7\ \ 8$ &
    $9\ 10\ 11$& $12\ 13$ & $14\ 15$  & $16\ 17$ & $18\ 19$& $20\ 21$& $22\ 23\ 24$  \\ \hline
	& & & & & & & & & &  \\ \hline
	${\tau_0}:=$ &  $e_1\ e_1\ e_1\ e_1$ &  $e_2\ e_2\ e_2\ e_2$ &  $e_3\ e_3\ e_3$ &  $e_4\ e_4$ & 
    $e_5\ e_5$& $e_6\ e_6$& $e_7\ e_7$& $e_8\ e_8$& $e_9\ e_9\ e_9$  \\ \hline
    ${\sigma_0}:=$ &  $n_2\ n_9\ n_7\ n_1$ &  $n_9\ n_8\ n_5\ n_1$ &  $n_7\ n_9\ n_6$ &  $n_7\ n_6$ & 
    $n_5\ n_1$& $n_7\ n_3$& $n_4\ n_2$& $n_3\ n_4$& $n_2\ n_4\ n_8$  \\ \hline
    
    & & & & & & & & & &  \\ \hline
   Step 1  &$1\ \sm\ \sm\ \sm$ & & & & & & & & &  \\ \hline
    Step 2  & $1\ \ 0\ \ 0\ \ 0$& $\sm\hspace{9mm} \sm$  &$\sm\sm\hspace{4mm} $ &$\sm\hspace{5mm} $ &$\hspace{5mm} \sm$ &$\sm\hspace{5mm} $ & & & &  \\ \hline
    Step 3  & &$0\ \ 0\ 0\ \ 1$  &$0\ 1\ \ 0$ & $1\ \ 0$ &$0\ \ 1$  &$1\ \ 0$ & &$\sm\hspace{5mm}$ &$\hspace{11mm}\sm$ &  \\ \hline
    Step 4 & &  & & &  & &$\sm\sm$ &$1\ 0$ &$0\ \ 0\ \ 1$ &  \\ \hline
     Step 5 & &  & & &  & &$1\ 0$ & & &  \\ \hline     
 & & & & & & & & & &  \\ \hline  
    $y:=$   &$1\ \ 0\ \ 0\ \ 0$ & $1\ \ 0\ \ 0\ \ 0$ &$ 0\ \ 1\ \ 0$ &$1\ \ 0$ & $ 0\ \ 1$ &$ 1\ \ 0$
    &$ 1\ \ 0$ &$ 1\ \ 0$ &$0\ \ 0\ \ 1$ &  \\ \hline\hline
      & & & & & & & & & &  \\ \hline
      
   ${\tau_0}:=$ &  $e_1\ e_1\ e_1\ e_1$ &  $e_2\ e_2\ e_2\ e_2$ &  $e_3\ e_3\ e_3$ &  $e_4\ e_4$ & 
    $e_5\ e_5$& $e_6\ e_6$& $e_7\ e_7$& $e_8\ e_8$& $e_9\ e_9\ e_9$  \\ \hline
    ${\sigma_1}:=$ &  $n_2\ n_{9}\ n_7\ n_1$ & $n_9\ n_{1}\ n_5\ n_1$ &  $n_7\ n_9\ n_{6}$ &  $n_7\ n_6$ &  $n_{5}\ n_1$& $n_3\ n_8$& $n_{4}\ n_{2}$& $n_3\ n_4$& $n_8\ n_8\ n_8$  \\ \hline
     & & & & & & & & & &  \\ \hline
        &$1\ \sm\ \sm\ \sm$ & & & & & & & & &  \\ \hline
     & $1\ \ 0\ \ 0\ \ 0$& $\  \sm\sm \hspace{5mm} \sm$  &$\sm\sm\hspace{4mm} $ &$\sm \hspace{5mm} $ &$\hspace{5mm} \sm$ & & & & &  \\ \hline
     & &$0\ \  0\ \ 0\ \ 1$  &$0\ \ 1\ \ 0$ & $1\ \ 0$ &$0\ \ 1$  & & & & &  \\ \hline
      & &  & &  & &$0_5\ 1_6$ &$1_2\ 0_1$ &$1_4\ 0_3$ &$0_7\ 0_7\ 1_8$ &  \\ \hline
 
        \end{tabular}

        {\sl Table 13: Attempts to find {\bf one} bitstring in a set of type $(e-row)\cap(n-row)$}

\vspace{3mm}
{\bf 9.2} Let us investigate a slight variation $\sigma_1$ of $\sigma_0$ (still in Table 13) where $\stackrel{\ra}{n_2}$ will not get calmed automatically. How to construct a bitstring $y'\in\tau_0\cap\sigma_1$ nevertheless? Our first step is identical to Step 1 of 9.1, while our second step is {\it almost} identical (up to the last smiley) to Step 2 of 9.1. However now there are no longer inviting positions on offer. In particular the wildcard $\stackrel{\ra}{n_2}$ is not yet calmed. We therefore put out the "n-fire" of position 19 by setting $y'_{19}:=0\ (=0_1)$. Because $\stackrel{\ra}{e_7}$ has length 2, this unfortunately triggers an "e-fire" on position 18. Putting it out by setting $y'_{18}:=1_2$ triggers a new n-fire, and so it goes on with $0_3,1_4,0_5,1_6$. Fortunately in the end 
$y'_{22}=y'_{23}:=0_7,\ y'_{24}:=1_8$ completes the definition of a bitstring $y'$ in $\tau_0\cap\sigma_1$. Once more we were lucky, yet an idea how to handle the general case
does not emerge. To just mention one issue, how to handle potential $2$'s in the original $2e$- or $2n$-row?
In Subsections 9.3 and 9.4 we tackle  matters  differently.

{\bf 9.3} It is natural to couple a  bipartite graph $B=B(\rho,\sigma)$ to any given $2e$-row $\rho$ and same length $2n$-row $\sigma$. Its two shores are made up, respectively by the wildcards $\stackrel{\ra}{e}$ of $\rho$ and the wildcards $\stackrel{\ra}{n}$ of $\sigma$.
By definition 
vertices $\stackrel{\ra}{e}$ and $\stackrel{\ra}{n}$ of $B$ are adjacent iff $pos(\stackrel{\ra}{e})\cap pos(\stackrel{\ra}{n})\neq\es$. Do matters simplify by looking at the 
connected components of $B$? In 9.1 the whole graph $B$ was connected and could indeed be traced in one go
(by repeatedly accepting inviting positions). In 9.2 the graph $B$ was also connected but could not be traced that easily.

Here in 9.3 we show that for special types of $\rho$ and $\sigma$ the behaviour of $B$ is more predictible. Namely $\rho$ and $\sigma$ must both be {\it lean} in the sense that all their wildcards have length 2. For instance $\rho',\sigma'$ in Table 14 are lean.

\begin{tabular}{l|c|c|c|c|c|c|c|c|c|c|c|c|c|}	
	&1 &2 &3 &4 &5 &6 &7 &8 &9 &10 &11 &12 &13    \\ \hline
	& & & & & & & & & & & & &  \\ \hline	
	$\rho'=$ &  2 &  $e_2$ &$e_1$ &2 & $e_4$& 2 &$e_3$&  $e_2$&  $e_4$ &  $e_3$ & 2 &  $e_1$ & 2    \\ \hline
    $\sigma'=$ &  $n_5$ &  $n_4$ &$n_2$ &$n_3$ & $n_2$& $n_3$ &$n_6$&  $n_4$&  $n_1$ &  $n_5$ & 2 &  $n_1$ &$n_6$    \\ \hline
    \end{tabular}

\vspace{3mm}
{\sl Table 14: Intersecting a lean 2e-row with a lean 2n-row}

As is true  for any two lean rows, each vertex of $B':=B(\rho',\sigma')$ has degree $\le 2$. This implies that each connected component is either an isolated vertex, a path, or a cycle.
Our $B'$, shown  in Fig. 3, has four connected components. The wildcard 
$\stackrel{\longrightarrow}{n_3}:=(n_3,n_3)$  constitutes an isolated component of $B'$ because $pos(\stackrel{\longrightarrow}{n_3})=\{4,6\}\s twos(\rho')$. If, say\footnote{Instead of (0,0), also (0,1) or (1,0) work, but not (1,1).}, we define $(y_4,y_6):=(0,0)$, then the bitstring $(y_4,y_6)$ "satisfies" the connected component $\big\{\stackrel{\longrightarrow}{n_3}\big\}$ (i.e. the corresponding formula $\ol{x_4}\vee\ol{x_6}$).

\begin{center}
   \includegraphics[scale=0.6]{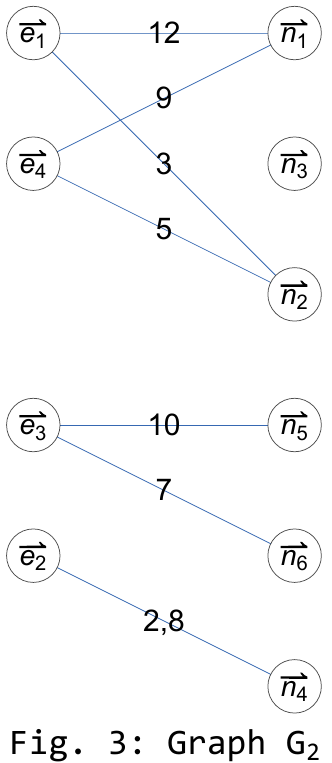} 
\end{center}

As to the connected component $\big\{\stackrel{\longrightarrow}{n_4},\stackrel{\longrightarrow}{e_2}\big\}$, it holds that
$pos(\stackrel{\longrightarrow}{n_4})=pos(\stackrel{\longrightarrow}{e_2})=\{2,8\}$. If we therefore put $(y_2,y_8):=(1,0)$ (or: (0,1)), then $(y_2,y_8)$ satisfies this connected component (which classifies as path, albeit one consisting of a single edge).

As to the connected component $\big\{\stackrel{\longrightarrow}{n_5},\stackrel{\longrightarrow}{e_3}
,\stackrel{\longrightarrow}{n_6}\big\}$, it is a path as well. One checks that 
$$pos(\stackrel{\longrightarrow}{n_5})=\{1,10\},\ pos(\stackrel{\longrightarrow}{e_3})=\{10,7\},\ pos(\stackrel{\longrightarrow}{n_6})=\{7,13\}.$$
Furthermore\footnote{Generally for each connected component which is a path $P$ of length $m\ge 2$ the following holds. Let $\alpha,\beta$ the two unique positions covered by $P$ which satisfy $\alpha,\beta\in twos(\rho)\cup twos(\sigma)$. If $m$ is even then either $\alpha,\beta\in twos(\rho)$ or $\alpha,\beta\in  twos(\sigma)$.
If $m$ is odd, then either ($\alpha\in twos(\rho),\beta\in twos(\sigma)$) or ($\beta\in twos(\rho),\alpha\in twos(\sigma)$).} $\{1,13\}\s twos(\rho')$. It follows that $(y_1,y_{10},y_7,y_{13}):=(0,1,0,1)$ (or: (1,0,1,0))  satisfies the connected component $\big\{\stackrel{\longrightarrow}{n_5},\stackrel{\longrightarrow}{e_3}
,\stackrel{\longrightarrow}{n_6}\big\}$.

The fourth connected component $\big\{\stackrel{\longrightarrow}{e_1},\stackrel{\longrightarrow}{n_2}
,\stackrel{\longrightarrow}{e_4},,\stackrel{\longrightarrow}{n_1}\big\}$ is a 4-cycle which is satisfied by the bitstring $(y_3,y_5,y_9,y_{12}):=(1,0,1,0)\ (or: (0,1,0,1))$. In a bipartite graph all cycles are even. Hence $(1,0,1,0)$ generalizes to $(1,0,1,0,...,1,0)$, and
$(0,1,0,1)$ generalizes to $(0,1,0,1,...,0,1)$. 

By merging the four constructed bitstrings one obtains a bitstring $(y_1,y_2,..,y_{13})$ that satisfies the whole of $B'$, i.e. lies in $\rho'\cap\sigma'$.

\vspace{4mm}

{\bf 9.4} The pleasant state of affairs for  lean rows suggests to somehow reduce the general case to the scenario in 9.3.  This eventually led to the inductive scheme in Lemma 4 below. 
First some definitions are in order. Let $\rho$ be a $2e$-row indexed by $1,2,...,m$.
Then $(Nu,Ei)$ is a {\it 01-injection} of $\rho$ if $Nu\uplus Ei\s\{1,...,m\}$ is such that $|pos(\stackrel{\ra}{e})\setminus Nu|\ge 2$ for all wildcards $\stackrel{\ra}{e}$ of $\rho$. Consider $\rho$ in Table 15 (and ignore $\sigma$ for the time being):

\begin{tabular}{l|c|c|c|c|c|c|c|c|c|c|c|c|c|c|c|c|c|c|c|c|}
&1 &2 &3 &4 &5 &6 &7 & &1 &2 &3 &4 &5 &6 &7   & &1 &3 &4 & 7   \\ \hline
& & & & & & & & & & & & & & &   & & & & &    \\ \hline
	$\rho:=$&$e_1$ &$e_1$ & $e_1$&$e_2$ &2 &$e_2$ &$e_2$ &$\ \Ra\ $ &$e_1$ &$\bf 0$ &$e_1$ &$e_2$ &$\bf 1$
    &$\bf 0$ &$e_2$  &$\ \Ra\ \rho':=$ &$e_1$ &$e_1$ &$e_2$&$e_2$\\ \hline
    $\sigma:=$&$n_3$ &$n_3$ & $n_1$&$n_1$ &$n_1$ &$n_2$ &$n_2$ &$\ \Ra\ $ &2 &$\bf 0$ &$n_1$ &$n_1$ &$\bf 1$ &$\bf 0$ &2  &$\ \Ra\ \sigma':=$ &2 &$n_1$ &$n_1$& 2\\ \hline
     \end{tabular}

     {\sl Table 15: Applying 01-injections}

Thus $(\{2,6\},\{5\})$ is a 01-injection of $\rho$ since $pos(\stackrel{\ra}{e_1})\setminus \{2,6\}=\{1,3\}$ and $pos(\stackrel{\ra}{e_2})\setminus \{2,6\}=\{4,7\}$, yet $(\{2,3\},\es)$ is not since
 $pos(\stackrel{\ra}{e_1})\setminus \{2,3\}=\{1\}$. By {\it applying} $(Nu,Ei):=(\{2,6\},\{5\})$
 to $\rho$ we get the $2e$-row $\rho'$ in Table 15. Formally, let $\tau$ be the 012-row defined by
 $zeros(\tau):=Nu, ones(\tau):=Ei$, and $twos(\tau):=(Nu\uplus Ei)^c=\{1,3,4,7\}$. Hence $\rho\cap\tau$
 is the top middle row in Table 15, and $\rho'$ is obtained by picking those components of $\rho\cap\tau$ that have indices in $(Nu\uplus Ei)^c$.
 Generally the application of 01-injections to $2e$-rows yields again $2e$-rows.
 (In contrast, applying the faulty $(\{2,3\},\es)$ to $\rho$  yields $\tau=(2,0,0,2,2,2,2)$, hence $\rho\cap\tau=(1,0,0,e_2,2,e_2,e_2)$, hence $\rho'=({\bf 1},e_2,2,e_2,e_2)$, which is {\it not} a 2e-row.)

Dually, let $\sigma$ be a $2n$-row indexed by $\{1,2,...,m\}$. Then $(Nu,Ei)$ is a {\it 01-injection} of $\sigma$ if $Nu\uplus Ei\s\{1,...,m\}$ is such that $|pos(\stackrel{\ra}{n})\setminus Ei|\ge 2$ for all wildcards $\stackrel{\ra}{n}$ of $\sigma$. Similarly to before one defines "applying" and argues that applying 01-injections to $2n$-rows $\sigma$ results in $2n$-rows $\sigma'$ of usually\footnote{Of course, applying 01-injections of type
$(Nu,\es)$ to $\sigma$ yields $\sigma'=\sigma$. Dually, applying $(\es, Ei)$ to 2e-rows has no effect.} shorter length.

The problem we face is that for given {\it (2e,2n)-pairs} $(\rho,\sigma)$ we seek 01-injections that  simultaneously apply to both $\rho$ and $\sigma$. For instance $(\{2,6\},\{5\})$ is such a 
{\it simultaneous (sim.)} 01-injection for the $(2e,2n)$-pair $(\rho,\sigma)$ in Table 15. Upon application one gets the $(2e,2n)$-pair $(\rho',\sigma')$ in Table 15.

{\bf Lemma 4: }{\it For each $(2e,2n)$-pair $(\rho,\sigma)$ there is a sim. 01-injection  $(Nu^*,Ei^*)$ with the following property. Upon applying $(Nu^*,Ei^*)$ to $(\rho,\sigma)$ the resulting
$(2e,2n)$-pair $(\rho^*,\sigma^*)$ is lean, i.e. all wildcards occuring in $\rho^*$ and $\sigma^*$ have length 2.}

{\it Proof.} We  induct on the common length $m$ of $\rho$ and $\sigma$. The claim being true for $m\le 2$, we henceforth assume that $m\ge 3$ and embark on a 3-fold case distinction.
In all cases $(\rho,\sigma)$ will be reduced to a shorter pair $(\rho',\sigma')$ by virtue of some sim. 01-injection $(Nu,Ei)$, which is {\it elementary} in the sense that $|Nu|,|Ei|\le 1$. By induction some sim. 01-injection $(Nu',Ei')$ carries $(\rho',\sigma')$
 to the claimed type $(\rho^*,\sigma^*)$. Evidently $(Nu^*,Ei^*):=(Nu\cup Nu',Ei\cup Ei')$ then carries $(\rho,\sigma)$ to  $(\rho^*,\sigma^*)$.

 {\bf Case 1:} Either $twos(\rho)\neq\es$ or $twos(\sigma)\neq\es$. {\it Subcase A:} There is $i\in twos(\rho)\cap twos(\sigma)$. Then applying say\footnote{Applying $(\es,\{i\})$ has the same effect.} $(Nu,Ei):=(\{i\},\es)$ to $(\rho,\sigma)$ yields a $(2e,2n)$-pair $(\rho',\sigma')$ whose sole difference to  $(\rho,\sigma)$ is one lost (common) 2-symbol at position $i$. {\it Subcase B:} $twos(\rho)\cap twos(\sigma)=\es\ and\ twos(\rho)\neq\es$.
 Then pick any $i\in twos(\rho)$ and put $(Nu,Ei):=(\{i\},\es)$. Applying $(Nu,Ei)$ to $(\rho,\sigma)$ yields a $(2e,2n)$-pair $(\rho',\sigma')$ with these properties: $\rho'$ results from $\rho$ by dropping a dont-care 2 on position $i$, and $\sigma'$ results from $\sigma$ by
 dropping the symbol $n$ on position $i$ of a wildcard $\stackrel{\ra}{n}$ and by turning all other components of $\stackrel{\ra}{n}$ to "2". {\it Subcase C:} $twos(\rho)\cap twos(\sigma)=\es\ and\ twos(\sigma)\neq\es$. Then take $(Nu,Ei):=(\es,\{i\})$ and proceed dually to Subcase B. (The Subcases B and C are not mutually exclusive. After repeated application of Subcases A,B,C one obtains two rows (call them again $\rho,\sigma$) with $twos(\rho)=twos(\sigma)=\es$.)

 {\bf Case 2:} $twos(\rho)=twos(\sigma)=\es$ and there is a wildcard $\stackrel{\ra}{e}$ of $\rho$ with
 $|pos(\stackrel{\ra}{e})|=k\ge 3$. Fix $i\in pos(\stackrel{\ra}{e})$ and put again $(Nu,Ei):=(\{i\},\es)$. Now the effect of applying $(Nu,Ei)$ to $\rho$ is different: the $e$-symbol on position $i$ gets killed and $\stackrel{\ra}{e}$ gives way\footnote{If we had $|pos(\stackrel{\ra}{e})|=2$ then $\stackrel{\ra}{e}$ would give way to a 1-bit, which contradicts the definition of a 01-injection. This goes to show that lean $(2e,2n)$-pairs need not allow 01-injections.} to a wildcard of length $k-1\ge 2$. The effect of applying $(Nu,Ei)$ to $\sigma$ is obvious if $i\in twos(\sigma)$, and otherwise the effect is the same as in Case 1.

 {\bf Case 3:} $twos(\rho)=twos(\sigma)=\es$ and all wildcards of $\rho$ have length 2, and there is a wildcard $\stackrel{\ra}{n}$ of $\sigma$ with
 $|pos(\stackrel{\ra}{n})|=k\ge 3$. This case is much dual to Case 2 and is left to the reader.

 If none of these  cases apply, then  $\rho$ and $\sigma$ are already lean (and without 2's). $\square$
 
{\bf 9.4.1} To illustrate, consider the $(2e,2n)$-pair $(\rho,\sigma)$ in Table 16 which coincides with $(\ol{\rho_3},\ol{\sigma_3})$ from 3.4. Applying twice Subcase B of Case 1 yields the $(2e,2n)$-pair $(\rho^*,\sigma^*)$ (cancel the bracketed 0's). Applying twice Subcase C of Case 1 yields the $(2e,2n)$-pair $(\rho^{**},\sigma^{**})$, where both $\rho^{**}:=(2,2,2,2)$ and $\sigma^{**}:=(n_2,n_2,n_4,n_4)$ are lean. Being a lean $(2e,2n)$-pair, $(\rho^{**},\sigma^{**})$ can be handled as in 9.3. One also sees ad hoc that e.g. $(1',0',0',1')\in \rho^{**}\cap \sigma^{**}$, and so $({\bf 0},1',{\bf 1},0',{\bf 0},0',1',{\bf 1})\in \rho\cap\sigma$.

\begin{tabular}{l|c|c|c|c|c|c|c|c|l}
\hline
	$\rho:=$ &  2 & $e_1$ &$e_1$ & $e_1$  & 2& $e_4$  & $e_4$ & $e_4$  \\ \hline
    $\sigma:=$ &  $n_3$ & $n_2$ &$2$ & $n_2$  & $n_3$& $n_4$  & $n_4$ & $n_3$  \\ \hline
    &&&&&&&&\\ \hline
    $\rho^*:=$ &  (0) & $e_1$ &$e_1$ & $e_1$  & (0)& $e_4$  & $e_4$ & $e_4$  \\ \hline
    $\sigma^*:=$ &  $(0)$ & $n_2$ &$2$ & $n_2$  & $(0)$& $n_4$  & $n_4$ & $2$  \\ \hline
    &&&&&&&&\\ \hline
    $\rho^{**}:=$ &  ({\bf 0}) & $2$ &$({\bf 1})$ & $2$  & ({\bf 0})& $2$  & $2$ & $({\bf 1})$  \\ \hline
    $\sigma^{**}:=$ &  $({\bf 0})$ & $n_2$ &$({\bf 1})$ & $n_2$  & $({\bf 0})$& $n_4$  & $n_4$ & $({\bf 1})$  \\ \hline
\end{tabular}

{\sl Table 16: Applying 01-injections to the (2e,2n)-pair from Subsection 3.4.}

{\bf 9.4.2} We are now in a position to tie together various loose ends and to prove Theorem 1 from Subsection 3.6.

{\bf Theorem 1:} {\it Let $f:\{0,1\}^m\to\{0,1\}$ be a Boolean function which is in CNF format and of type
{\tt DisjointPossitive$\hspace{2pt} \wedge\hspace{2pt} $DisjointNeggative}. Then the satisfiability of $f$ can be tested in $O(m^2)$ time.}

{\it Proof.} As shown in 3.6, upon investing $O(m^2)$ time one can achieve the following. Either prove that $f$ is insatisfiable (because of trivial reasons, as defined in 3.3), or obtain a 2e-row $\ol{\rho_t}$ and a 2n-row $\ol{\sigma_t}$ with these properties.  They have the same length ($\ol{m}\le m$) and it holds that $\ol{\rho_t}\cap \ol{\sigma_t}\neq\es$ iff $f$ is satisfiable. In the remainder we prove that $\ol{\rho_t}\cap \ol{\sigma_t}\neq\es$ {\it always} happens. Thus we will
pinpoint some bitstring $y\in\ol{\rho_t}\cap \ol{\sigma_t}$.

To begin with, we can assume that both $\ol{\rho_t}$ and $\ol{\sigma_t}$ are indexed by $1,2,...,\ol{m}$.
According to Lemma 4 there is some simultaneous 01-injection $(Nu^*,Ei^*)$ which, applied to  $(\ol{\rho_t},\ol{\sigma_t})$, yields  a lean $(2e,2n)$-pair $(\rho',\sigma')$. The common index set of $\rho'$ and $\sigma'$ is $\{1,2,...,\ol{m}\}\setminus (Nu\uplus Ei)$. According to Subsection 9.3 there is a bitstring $y'\in \rho'\cap\sigma'$. Define a length $\ol{m}$ bitstring $y$ as follows.
For each $i\in\{1,...,\ol{m}\}$  put $y_i:=0$ if $i\in Nu$, put $y_i:=1$ if $i\in Ei$, and put $y_i:=y'_i$ otherwise. It is clear that $y\in\ol{\rho_t}\cap \ol{\sigma_t}$. $\square$

{\bf 9.5} Here we show how Theorem 1 can be used to find, in output-polynomial time {\it all} bitstrings in the intersection of a 2e-row $\rho$ with a same length 2n-row $\sigma$. 
To fix ideas, consider the e-row $\rho$ in Table 17 and the n-row $\sigma$ below. (Don't-care 2's will come up later, automatically).

\begin{tabular}{l|c|c|c||c|c||c|c|c||c|c|l}
	& 1 & 2 & 3& 4 & 5  & 6   & 7& 8& 9& 10 & \\ \hline
	& & & & & & & & & & & \\ \hline
	${\rho}:=$ &  $e_1$ & $e_2$ & $e_4$ & $e_1$ & $e_3$ & $e_3$& $e_3$&  $e_4$&  $e_4$ & $e_2$ & pending $n_1$\\ \hline
    $\sigma:=$ &  $n_1$ & $n_1$ & $n_1$ & $n_2$ & $n_2$ & $n_3$& $n_3$&  $n_3$&  $n_4$ & $n_4$ & \\ \hline
    & & & & & & & & & & & \\ \hline
    ${\rho_1}:=$ &  $\bf 0_1$ & $\bf e_2$& $\bf e_4$ & $1_2$ & $0_3$  & $e_3$& $e_3$&  $e_4$& 
    $e_4$ & $e_2$ &  p. $n_3$ \\ \hline
     ${\rho_2}:=$ &  $\bf 1$ & $\bf 0_1$& $\bf e_4$ & $2$ & $e_3$  & $e_3$& $e_3$&  $e_4$& 
    $0_3$ & $1_2$ &  p. $n_2$ \\ \hline
     ${\rho_3}:=$ &  $\bf 1$ & $\bf 1$& $\bf 0$ & $2$ & $e_3$  & $e_3$& $e_3$&  $e_4$& 
    $e_4$ & $2$ &  p. $n_2$ \\ \hline
    
     & & & & & & & & & & & \\ \hline
      ${\rho_{1,1}}:=$ &  $0$ & $e_2$& $e_4$ & $ 1$ & $0$  & $\bf 0$& $\bf 1$&  $\bf e_4$& 
       $e_4$ & $e_2$ &  p. $n_4\ \Ra\ 2\ f.r.\ of\ card=9+4$ \\ \hline
       ${\rho_{1,2}}:=$ &  $0$ & $e_2$& $e_4$ & $ 1$ & $0$  & $\bf 1$& $\bf 0$&  $\bf e_4$& 
       $e_4$ & $e_2$ &  p. $n_4\ \Ra\ 2\ f.r.\ of\ card=9+4$ \\ \hline
      ${\rho_{1,3}}:=$ &  $0$ & $e_2$& $e_4$ & $ 1$ & $0$  & $\bf 1$& $\bf 1$&  $\bf 0$& 
       $e_4$ & $e_2$ &  p. $n_4\ \Ra\ 2\ f.r.\ of\ card=3+2$ \\ \hline
 ${\rho_2}:=$ &  $1$ & $0$& $ e_4$ & $2$ & $e_3$  & $e_3$& $e_3$&  $e_4$& 
    $0$ & $1$ &  p. $n_2$ \\ \hline
     ${\rho_3}:=$ &  $1$ & $1$& $ 0$ & $2$ & $e_3$  & $e_3$& $e_3$&  $e_4$& 
    $e_4$ & $2$ &  p. $n_2$ \\ \hline   
     & & & & & & & & & & & \\ \hline
 ${\rho_{2,1}}:=$ &  $1$ & $0$& $ e_4$ & $\bf 0$ & $\bf e_3$  & $e_3$& $e_3$&  $e_4$& 
    $0$ & $1$ &  p. $n_3$ \\ \hline
${\rho_{2,2}}:=$ &  $1$ & $0$& $ e_4$ & $\bf 1$ & $\bf 0$  & $e_3$& $e_3$&  $e_4$& 
    $0$ & $1$ &  p. $n_3\Ra\ 3\ f.r.\ of\ card= 3+3+1$ \\ \hline
     ${\rho_3}:=$ &  $1$ & $1$& $ 0$ & $2$ & $e_3$  & $e_3$& $e_3$&  $e_4$& 
    $e_4$ & $2$ &  p. $n_2\Ra\ 10\ f.r.\ of\ card=...=31$ \\ \hline 
    
        & & & & & & & & & & & \\ \hline

        ${\rho_{2,1,1}}:=$ &  $1$ & $0$& $ e_4$ & $0$ & $ e_3$  & $\bf 0$& $\bf e_3$&  $\bf e_4$& 
    $0$ & $1$ &  final,\ card=9 \\ \hline
    ${\rho_{2,1,2}}:=$ &  $1$ & $0$& $ e_4$ & $0$ & $2$  & $\bf 1$& $\bf 0$&  $\bf e_4$& 
    $0$ & $1$ &  final,\ card=6 \\ \hline
     ${\rho_{2,1,3}}:=$ &  $1$ & $0$& $ 1$ & $0$ & $2$  & $\bf 1$& $\bf 1$&  $\bf 0$& 
    $0$ & $1$ &  final,\ card=2 \\ \hline
     \end{tabular}

      {\sl Table 17: Calculating {\it all} bitstrings in sets of type $(2e-row)\cap(2n-row)$}

Recall from 2.2 that the following is an instance of the general behaviour of Abraham 0-Flags:
$$(25)\quad (n_1,n_1,n_1)=(0,2,2)\uplus(1,0,2)\uplus(1,1,0)$$
Accordingly, with $(n_1,n_1,n_1)$ coming from $\sigma$, we conclude 
$$\rho\cap({\bf n_1,n_1,n_1},2,2,2,2,2)=\big(\rho\cap({\bf 0,2,2},2,2,2,2,2,2,2)\big)
\ \uplus\  \big(\rho\cap({\bf 1,0,2},2,2,2,2,2,2,2)\big)$$
$$\uplus \big(\rho\cap({\bf 1,1,0},2,2,2,2,2,2,2)\big)=:\ \rho_1\uplus\rho_2\uplus\rho_3,$$

where $\rho_1,\rho_2,\rho_3$ are defined in Table 17.
As in 4.2 we call  them the canddate sons of $\rho$.
The  components $0_1,1_2,0_3$ of $\rho_1$ indicate 0,1-propagation (as on the bottom of Table 13). Likewise for $\rho_2.$
Using the terminology of 9.1, all bitstrings in $\rho_1\uplus\rho_2\uplus\rho_3$ calm $\stackrel{\rightarrow}{n_1}$. Incidently $\rho_1$ also calms $\stackrel{\rightarrow}{n_2}$, and so $\stackrel{\rightarrow}{n_3}$ is {\it pending}. For $\rho_2,\rho_3$ the imposition of
 $\stackrel{\rightarrow}{n_2}$ is pending.

{\bf 9.5.1} Before continuing in 9.5.2 we need to pause and ask: Concerning general starter rows $\rho$ and $\sigma$ (of type 2e and 2n), is it true that while imposing $n$-wildcards all arising 012e-rows $\ol{\rho}$ stay\footnote{In accordance with the definition of Sec. 4.2 here "feasible" means that $\ol{\rho}\cap\sigma\neq\es$.} feasible?  Not necessarily, but infeasible rows can be detected fast and eliminated. Namely, consider  $\ol{\rho}:=\rho_1$ in Table 17. From $\rho\cap\sigma\neq\es$ (Thm.1) does not follow that $\ol{\rho}\cap\sigma\neq\es$ (since $\ol{\rho}\s\rho$). In order to decide $\ol{\rho}\cap\sigma\stackrel{?}{=}\es$, upon applying 0,1-propagation (if necessary) one either finds in $O(m^2)$ time that $\ol{\rho}\cap\sigma{=}\es$ by trivial reasons, or that
$\ol{\rho}\cap\sigma\neq\es$ by Theorem 1. In the former case $\ol{\rho}$ is infeasible and gets deleted. In the latter case $\ol{\rho}$ is feasible, and {\it at least one} of its candidate sons will remain feasible (since the union of all its candidate sons contains $\ol{\rho}\cap\sigma$).

{\bf 9.5.2} We now resume (without feasibility tests, for brevity) the imposition of the pending $n$-wildcards in Table 17. Recall that always the top row of the LIFO stack, here $\rho_1$, comes first. Upon using again an Abraham 0-flag it follows that now $\stackrel{\ra}{n_4}$  is pending in all of $\rho_{1,1},\rho_{1,2},\rho_{1,3}$.
In order to impose $\stackrel{\ra}{n_4}$ upon $\rho_{1,1}$ one has to set (Abraham $2\times 2$ flag) its last two components to
$(e_4,e_2):=(0,e_2)$, respectively  $(e_4,e_2):=(1,0)$. (The latter choice forces the first $e_2$ to become $1$).
We see that $\rho_{1,1}$ gives way to two final rows of cardinalities $3\cdot3$ and $2\cdot 2$. Similarly $\rho_{1,2}$ gives way to two final rows whose cardinalities sum up as $9+4$.
For $\rho_{1,3}$ we get $3+2$. Upon removing the six final rows from the LIFO stack, its rows are now $\rho_2$  and $\rho_3$.
The top row $\rho_2$ gives way to $\rho_{2,1}$ and $\rho_{2,2}$ which both have $\stackrel{\ra}{n_3}$ pending. While we detail  how $\rho_{2,1}$ gives way to 3 final rows, handling $\rho_{2,2}$ is left to the reader. Ditto for $\rho_3$. To summarize,
$\rho\cap\sigma$ contains 86 bitstrings which get packaged into 22 disjoint 012e-rows.

{\bf Theorem 5:} {\it Let $f:\{0,1\}^m\to\{0,1\}$ be a Boolean function which is in CNF format and of type {\tt DisjointPossitive$\hspace{2pt} \wedge\hspace{2pt} $DisjointNeggative}. Then $Mod(f)$ can be represented as disjoint union of $R$ many 012e-rows in time $O(Rm^3)$.}

{\it Proof.}  The algorithm outlined in Table 17 involves a so called  {\it row-splitting mechanism}; this rather abstract concept is fully defined in Section 8 of [W1]. If some Boolean function $\varphi$ admits such a row-splitting mechanism, then [W1, Thm.1] one can represent $Mod(\varphi)$ as disjoint union of $R$ multivalued rows in time $O(Rh(d+s))$. Translated to our scenario the parameter
$h$ is the number of $n$-wildcards in the $2n$-row $\sigma$ derived from {\tt DisjointNeggative}. (Recall from 3.5 that calculating $\sigma$ costs $O(m^2)$.) Further $s$ is the time it takes to compute the candidate sons of a fixed LIFO top row and to discard the infeasible ones among them. As seen in 9.5.1 here $s$ is $O(m^2)$. Finally $d$ is the time
it takes to compute which $n$-wildcard is pending in a candidate son. It is easy to see that also $d=O(m^2)$. It hence follows that $O(Rh(d+s))=O(Rm(m^2+m^2)=O(Rm^3)$. $\square$

In the proof above the $n$-wildcards of $\sigma$ were imposed one after the other upon the 2e- row $\rho$. Dually one could impose the $e$-wildcards upon the 2n-row $\sigma$. In practise one would impose the type of wildcard of which there are fewer\footnote{This is akin to Table 12 where some of the set-systems $\rho_i\cap\sigma_j$ could more easily be rewritten (ad hoc) as 012e-rows, and others as 012n-rows.}.

Because presumably not all readers know [W1,Thm.1], and by other reasons (e.g. verifying the claim about 86), we offer an alternative
proof of Theorem 5 in 9.7. Specifically, it will settle the two questions left open in  3.8.2.

{\bf 9.6} Sections 4 to 7 showed various applications of enumerating all bitstrings in sets $\rho\cap\sigma$ of type $(2e-row)\cap(2n-row)$. 
Recall from 5.1.2 the concept of a $g$-wildcard. Let now $\theta\cap\sigma$ be of type $({\bf 2g}-row)\cap(2n-row)$. This variant has the same complexity as Theorem 5 and has several applications as well (work in progress). In the present article the
alternative proof of Theorem 5 in 9.7 relies on the specific $({2g}-row)\cap(2n-row)$ type calculation carried out in 9.6.3.

{\bf 9.6.1} In order to show that $\theta\cap\sigma\neq\es$ in the first place,
 let $\rho$ be the 2e-row obtained from $\theta$ by replacing each $g$-wildcard by a $e$-wildcard
 of the same length. By Theorem 1 there is a bitstring $y\in\rho\cap\sigma$. By replacing suitable (yet not uniquely determined) 1's of $y$ with 0's one obtains a bitstring $y'$ such that $|ones(y')\cap pos(\stackrel{\ra}{e})|=1$ for all wildcards $\stackrel{\ra}{e}$ of $\rho$. Hence $y'\in \theta$, and  $y'\in \sigma$ since $zeros(y')\supseteq zeros(y)$.

{\bf 9.6.2} To fix ideas, let $\sigma$ be as in Table 17 and let $\theta$ be the $g$-row obtained from the $e$-row $\rho$ in Table 17 by replacing each $e$-wildcard with a same length $g$-wildcard.
Both rows are rendered again in Table 18.
We now show that enumerating $\theta\cap\sigma$ works considerably faster than enumerating $\rho\cap\sigma$. For starters, observe that on positions 6 and 7 we have $(g_3,g_3)$ and $(n_3,n_3)$. Each instantiation of the whole wildcard $(g_3,g_3,g_3)$ features two 0's and one 1. Since necessarily there is a 0 among its last two components, {\it each} bitstring $y\in\theta$ satisfies $\stackrel{\ra}{n_3}$. It thus remains to impose $\stackrel{\ra}{n_1},\stackrel{\ra}{n_2},\stackrel{\ra}{n_4}$ upon $\theta$.

        \begin{tabular}{l|c|c|c|c|c|c|c|c|c|c|l}
        & 1 & 2 & 3& 4 & 5  & 6   & 7& 8& 9& 10 & \\ \hline
        & & & & & & & & & & & \\ \hline
${\theta}:=$ &  $g_1$ & $g_2$ & $g_4$ & $g_1$ & $g_3$ & $g_3$& $g_3$&  $g_4$&  $g_4$ & $g_2$ & \\ \hline
 $\sigma:=$ &  $n_1$ & $n_1$ & $n_1$ & $n_2$ & $n_2$ & $n_3$& $n_3$&  $n_3$&  $n_4$ & $n_4$ & \\ \hline
    & & & & & & & & & & & \\ \hline
${\theta_1}:=$ &  $\bf 0_1$ & $\bf g_2$ & $\bf g_4$ & $1_2$ & $0_3$ & $g_3$& $g_3$&  $g_4$&  $g_4$ & $g_2$ & $pending\ n_4$\\ \hline
${\theta_2}:=$ &  $\bf 1$ & $\bf 0_1$ & $\bf g_4$ & $0$ & $g_3$ & $g_3$& $g_3$&  $g_4$&  $0_3$ & $1_2$ & final\\ \hline
${\theta_3}:=$ &  $\bf 1$ & $\bf 1_1$ & $\bf 0$ & $0$ & $g_3$ & $g_3$& $g_3$&  $g_4$&  $g_4$ & $0_2$ & final\\ \hline
 & & & & & & & & & & & \\ \hline
${\theta_{1,1}}:=$ &  $0$ & $ g_2$ & $ g_4$ & $1$ & $0$ & $g_3$& $g_3$&  $g_4$&  $\bf 0$ & $\bf g_2$ & final\\ \hline
${\theta_{1,2}}:=$ &  $0$ & $1$ & $0$ & $1$ & $0$ & $g_3$& $g_3$&  $0$&  $\bf 1$ & $\bf 0$ & final\\ \hline
 & & & & & & & & & & & \\ \hline\hline
  & & & & & & & & & & & \\ \hline
      
        ${\tau}:=$ &  $g_1$ & $g_1$ & $g_1$ & $g_2$ & $g_2$ & $g_3$& $g_3$&  $g_3$&  $g_4$ & $g_4$ & \\ \hline
    $\sigma_{aux}:=$ &  $n$ & $n'$ & $n''$ & $n$ & $2$ & $2$& $2$&  $n''$&  $n''$ & $n'$ & \\ \hline
     & & & & & & & & & & & \\ \hline
     ${\tau_1}:=$ &  $\bf 0$ & $g_1$ & $g_1$ & $\bf g_2$ & $g_2$ & $g_3$& $g_3$&  $g_3$
     &  $g_4$ & $g_4$ & pending $n'$\\ \hline
     ${\tau_2}:=$ &  $\bf 1$ & $0$ & $0$ & $\bf 0$ & $1$ & $g_3$& $g_3$&  $g_3$&  $g_4$ & $g_4$ & final, card=6\\ \hline
      & & & & & & & & & & & \\ \hline
       ${\tau_{1,1}}:=$ &  $0$ & $\bf 0$ & $1$ & $g_2$ & $g_2$ & $g_3$& $g_3$&  $g_3$
       &  $g_4$ & $\bf g_4$ & pending $n''$\\ \hline
     ${\tau_{1,2}}:=$ &  $0$ & $\bf 1$ & $0$ & $g_2$ & $g_2$ & $g_3$& $g_3$&  $g_3$&  $1$ & $\bf 0$ & final, card=6\\ \hline
      & & & & & & & & & & & \\ \hline
     $\tau_{1,1,1}:=$ &  $0$ & $0$ & $1$ & $g_2$ & $g_2$ & $g_3$& $g_3$&  $\bf 0$&  $\bf g_4$ & $g_4$ &final, card=8\\ \hline
      $\tau_{1,1,2}:=$ &  $0$ & $0$ & $1$ & $g_2$ & $g_2$ & $0$& $0$&  $\bf 1$&  $\bf 0$ & $1$ & final, card=2\\ \hline   
    \end{tabular}

    {\sl Table 18: Calculating {\it all} bitstrings in sets of type $(2g-row)\cap(2n-row)$}

Upon raising a $3\times 3$ Abraham 0-Flag the row $\theta$ gives way to $\theta_1,\theta_2,\theta_3$, all of which by construction calm $\stackrel{\ra}{n_1}$. All three  also happen to calm  $\stackrel{\ra}{n_2}$.
Furthermore $\theta_2,\theta_3$ happen to calm $\stackrel{\ra}{n_4}$, and hence are final. Therefore it remains to impose $\stackrel{\ra}{n_4}$ upon $\theta_1$. This results in the
 final rows $\theta_{1,1},\theta_{1,2}$.

{\bf 9.6.3}  We spell out a second example of this kind because it will be crucial in the alternative proof of Theorem 5. Thus consider $\tau$ and $\sigma_{aux}$ on the bottom of Table 18. Using the order of imposition  $n,n',n''$ yields
$\tau\cap\sigma_{aux}=\tau_2\uplus\tau_{1,2}\uplus\tau_{1,1,1}\uplus\tau_{1,1,2}$. In particular $|\tau\cap\sigma_{aux}|=22$.
(The mere cardinality is predictable via inclusion-exclusion, i.e. $36-6-6-2+0+0+0-0=22$. However, as usual inclusion-exclusion does not tell us how to list the counted objects, let alone listing them in compressed manner.)

{\bf 9.7} Let $(\rho,\sigma)$ be a $(2e,2n)$-pair. In 9.5 we  enumerated $\rho\cap\sigma$
by imposing the $n$-wildcards of $\sigma$ one-by-one upon $\rho$. Here we carry out the alternative
approach that we glimpsed in 3.8.
Namely, for the same $\sigma$ (repeated in Table 19) we write $\sigma=\sigma_{1,4,6,9}\uplus\sigma_{1,4,6,10}\uplus\cdots\uplus\sigma_{3,5,8,10}$ with 36 natural 012-rows $\sigma_{i,j,k,\ell}$ (details in 9.7.1). It then follows from distributivity that
$$(26)\quad\rho\cap\sigma=(\rho\cap\sigma_{1,4,6,9})\uplus (\rho\cap\sigma_{1,4,6,10})\uplus\cdots\uplus (\rho\cap\sigma_{3,5,8,10})$$
Each set of bitstrings $\rho\cap\sigma_{i,j,k,\ell}$ is either empty (by trivial reasons) or can be expressed as 012e-row (as in Table 4). Trouble is (9.7.2), in order to achieve output-polynomial time we must not stumble upon  quadruples $(i,j,k,\ell)$ which 
are {\it bad} in  that $\rho\cap\sigma_{i,j,k,\ell}$  is empty.

{\bf 9.7.1} With the wildcards $\stackrel{\rightarrow}{n_1},\stackrel{\rightarrow}{n_2},\stackrel{\rightarrow}{n_3},\stackrel{\rightarrow}{n_4}$ of $\sigma$ we associate certain Abraham 0-Flags (Figure 4) whose rows are labeled by the (subscripted) letters $\alpha,\beta,\gamma,\delta$.
Specifically:

\begin{itemize}
    \item[(27)] $(n_1,n_1,n_1)=(0,2,2)\uplus(1,0,2)\uplus(1,1,0)\ =:\ \alpha_1\uplus\alpha_2\uplus\alpha_3$
     \item[] $(n_2,n_2)=(0,2)\uplus(1,0)\ =:\ \beta_4\uplus\beta_5$
     \item[] $(n_3,n_3,n_3)=(0,2,2)\uplus(1,0,2)\uplus(1,1,0)\ =:\ \gamma_6\uplus\gamma_7\uplus\gamma_8$
        \item[] $(n_4,n_4)=(0,2)\uplus(1,0)\ =:\ \delta_9\uplus\delta_{10}$
\end{itemize}

\begin{center}
   \includegraphics[scale=1.2]{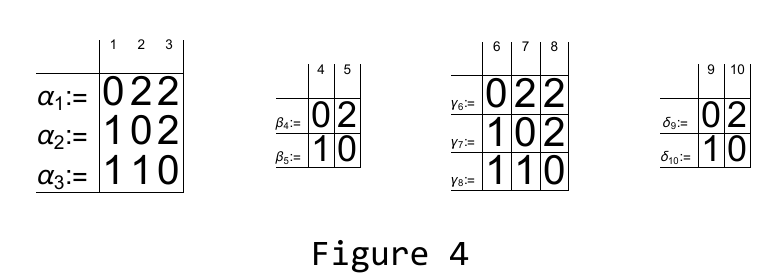} 
\end{center}

A hitting set $X$ w.r.t. a hypergraph $\HH $ is called {\it exact} if $|H\cap X|=1$ for all $H\in \HH $.
Hence the exact hitting sets (EHSes) $\{i,j,k,\ell\}$ of the hypergraph $\HH (\sigma):=\{\{1,2,3\},\{4,5\},\{6,7,8\},\{9,10\}\}$ bijectively\footnote{What about the fact that $\{2,5,7,9\}=\{9,2,7,5\}$, but $(2,5,7,9)\neq(9,2,7,5)$? This  does not affect  bijectivity since
$(9,2,7,5)\not\in\{1,2,3\}\times\{4,5\}\times\{6,7,8\}\times\{9,10\}$.} match the $3\cdot 2\cdot 3\cdot2=36$
quadruples $(i,j,k,\ell)\in \{1,2,3\}\times\{4,5\}\times\{6,7,8\}\times\{9,10\}$. The benefit of switching from quadruples to EHSes is that the latter can be modelled as the members of this g-row:
$$(28)\quad \tau:=(g_1,g_1,g_1,\ g_2,g_2,\ g_3,g_3,g_3,\ g_4,g_4)$$
If henceforth we speak of quadruples, think of them as members of $\tau$.
Each quadruple $(i,j,k,\ell)$  yields a 012-row $\sigma_{i,j,k,\ell}$ contained in $\sigma$.
For instance $(1,5,6,9)$ yields \\ $\sigma_{1,5,6,9}:=(0,2,2,\ 1,0,\ 0,2,2,\ 0,2)$, which arises by concatenating the 012-rows
$\alpha_1,\beta_5,\gamma_6,\delta_{9}$. It is clear that distinct 012-rows of this type are disjoint and that their union equals $\sigma$.

 The row $\sigma_{3,4,8,9}=(1,1,{\bf 0},\ 0,2,\ 1,1,{\bf 0},\ ,{\bf 0},2)$ has (among other such rows) the property that
$\rho\cap\sigma_{3,4,8,9}=\es$ since $pos(\stackrel{\ra}{e_4})=\{3,8,9\}\s zeros(\sigma_{3,4,8,9})$. In order to discard  {\it bad} quadruples like $\{3,4,8,9\}$ in advance, first observe that for  any\footnote{Here it is irrelevant whether or not $\sigma^*$ is contained in $\sigma$.} 012-row $\sigma^*$ of length 10 
an empty intersection $\rho\cap\sigma^*$ is {\it only} possible if $zeros(\sigma^*)$ contains at least one of $pos(e_1),pos(e_2),pos(e_3),pos(e_4)$. That's because
by (the proof of) Theorem 1 empty intersections are caused by trivial reasons and because
 0-1 clashes are ruled out in the present scenario ($\rho$ being a $2e$-row). Interestingly, {\it each} 012-row $\sigma_{i,j,k,\ell}$ satisfies $pos(e_3)\not\s zeros(\sigma_{i,j,k,\ell})$. Indeed, $pos(e_3)=\{5,6,7\}\s zeros(\sigma_{i,j,k,\ell})$ implies that $\sigma_{i,j,k,\ell}$ contains the pieces
 $\beta_5,\gamma_6,\gamma_7$ (since all 0's in the Abraham 0-Flags of Fig.4 occur in the diagonals). This contradicts the fact that $\{6,7\}\not\s\{i,j,k,\ell\}$ (recall that $\{i,j,k,\ell\}$ is an {\it exact} hitting set of $\HH(\sigma)$).

 {\bf 9.7.2} More systematically, which quadruples $\{i,j,k,\ell\}\in\tau$ yield  bad rows $\sigma_{i,j,k,\ell}$? For instance all rows $\sigma_{1,4,k,\ell}$ are bad because they contain the pieces $\alpha_1,\beta_4$, and so $pos(e_1)=\{1,4\}\s zeros(\sigma_{1,4,k,\ell})$ implies $\rho\cap\sigma_{1,4,k,\ell}=\es$. Likewise quadruples of type $\{2,j,k,10\}$ and $\{3,j,8,9\}$ are bad; and there are no other types of bad quadruples. In other words, $\{i,j,k,\ell\}\in\tau$ is {\it non-bad} iff it is a noncover of the hypergraph $\{\{1,4\},\{2,10\},\{3,8,9\}\}$. We therefore seek the members of $\tau\cap\sigma_{aux}$. Here $\tau$ is from (28) and $\sigma_{aux}:=(n,n',n'',n,2,2,2,n'',n'',n')$ is the {\it auxiliary} row coupled to $\sigma$.

\begin{tabular}{l|c|c|c||c|c||c|c|c||c|c|l}
	& 1 & 2 & 3& 4 & 5  & 6   & 7& 8& 9 & 10 & \\ \hline
    ${\rho}:=$ &  $e_1$ & $e_2$ & $e_4$ & $e_1$ & $e_3$& $e_3$& $e_3$&  $e_4$& $e_4$ & $e_2$ \\ \hline
     ${\sigma}:=$ &  $n_1$ & $n_1$ & $n_1$ & $n_2$ & $n_2$& $n_3$& $n_3$&  $n_3$& $n_4$ & $n_4$ \\ \hline
    &  &  & &  &   &    & & &  &  & \\ \hline
    $\sigma_{1,5,6,9}:=$ &  $0$ & $2$ & $2$ & $1$ & $0$& $0$& $2$&  $2$&
    $\bf 0$ & $\bf 2$ & where  $\{1,5,6,9\}\in\tau_2$\\ \hline
    {\bf 1.} $\rho\cap\sigma_{1,5,6,9}=$ &  $0$ & $e_2$ & $e_4$ & $1$ & $0$& $0$& $1$&  $e_4$&
        $0$ & $e_2$ &    cardinality=9\\ \hline 
        
 &  &  & &  &   &    & & &  &  & \\ \hline
        $\sigma_{1,5,6,10}:=$ &  $0$ & $2$ & $2$ & $1$ & $0$& $0$& $2$&  $2$&
    $\bf 1$ & $\bf 0$ & where  $\{1,5,6,10\}\in\tau_2$\\ \hline
   {\bf 2.}  $\rho\cap\sigma_{1,5,6,10}=$ &  $0$ & $1$ & $2$ & $1$ & $0$& $0$& $1$&  $2$&
        $1$ & $0$ & cardinality=4\\ \hline
        
 &  &  & &  &   &    & & &  &  & \\ \hline
  $\sigma_{2,4,6,9}:=$ &  $1$ & $0$ & $2$ & $0$ & $2$& $0$& $2$&  $2$&
    $0$ & $2$ & where  $\{2,4,6,9\}\in\tau_{1,2}$\\ \hline
   {\bf 3.}  $\rho\cap\sigma_{2,4,6,9}=$ &  $1$ & $0$ & $e_4$ & $0$ & $e_3$& $0$& $e_3$&  $e_4$& 
        $0$ & $1$ & cardinality=9\\ \hline
        
& & & & & & & & & & & \\ \hline
 $\sigma_{3,4,7,9}:=$ &  $1$ & $1$ & $0$ & $0$ & $2$& $1$& $0$&  $2$&
    $0$ & $2$ & where  $\{3,4,7,9\}\in\tau_{1,1,1}$\\ \hline
   {\bf 4.}  $\rho\cap\sigma_{3,4,7,9}=$ &  $1$ & $1$ & $0$ & $0$ & $2$& $1$& $0$&  $1$& 
        $0$ & $2$ & cardinality=4\\ \hline

        & & & & & & & & & & & \\ \hline
 $\sigma_{3,4,8,10}:=$ &  $1$ & $1$ & $0$ & $0$ & $2$& $1$& $1$&  $0$&
    $1$ & $0$ & where  $\{3,4,8,10\}\in\tau_{1,1,2}$\\ \hline
  {\bf 5.}   $\rho\cap\sigma_{3,4,8,10}:=$ &  $1$ & $1$ & $0$ & $0$ & $2$& $1$& $1$&  $0$&
    $1$ & $0$ & cardinality=2\\ \hline
    \end{tabular}

    {\sl Table 19: Five among 22 non-bad 012e-rows constituting $\rho\cap\sigma$}

The rows $\tau$ and $\sigma_{aux}$ conveniently happen to coincide with the same name rows in Table 18, where it was shown that $\tau\cap\sigma_{aux}=\tau_2\uplus\tau_{1,2}\uplus\tau_{1,1,1}\uplus\tau_{1,1,2}$. The latter were 01g-rows of cardinalities 6,6,8,2.
 In the present context the 22 good quadruples $\{i,j,k,\ell\}\in\tau\cap\sigma_{aux}$ yield exactly the 22 non-bad 012e-rows $\sigma_{i,j,k,\ell}$.  Table 19 displays five\footnote{The entusiastic reader will hasten to calculate all 22 of them and to confirm that their cardinalities add up to 86 (which coincides with the value obtained in 9.5).} random rows $\sigma_{i,j,k,\ell}\s\rho\cap\sigma$. For instance, the fact that $(1,5,6,9)$ differs
 from $(1,5,6,10)$ only in the last component, is reflected by $\sigma_{1,5,6,9}$ and $\sigma_{1,5,6,10}$   only differing in the last two components. (Nevertheless $\rho\cap\sigma_{1,5,6,9}$ and $\rho\cap \sigma_{1,5,6,10}$ differ in {\it five} components.)

\vspace{2cm}

All data is available upon request, Marcel Wild June-23, 2026

\section{References}

\begin{itemize}

\item[G] E. Mark Gold, Complexity of Automaton Identification from Given Data, Information and Control 37 (1978) 302-320.
\item[GJ] M.R. Garey, D.S. Johnson, Computers and intractability: A guide to the theory of NP-completeness, Freemann and Company, 1979.
\item[GV] A. Gainer-Dewar, P. Vera-Licona, The minimal hitting set generation problem: algorithms and computation, SIAM J. Discrete Math. 31 (2017) 63-100.
\item[HK] M.J.H. Heule, O. Kullmann, The science of brute force, Communications of the ACM 60 (2017) 70-79.

\item[MA] T. Mori, T. Akutsu, Attractor detection and enumeration algorithms for Boolean networks, Computational and Structural Biotechnology Journal
20 (2022) 2512-2520.
\item[MTY] K. Makino, S.Tamaki, M.Yamamoto, Derandomizing the HSSW algorithm for 3-SAT. Algorithmica 2013;67(2):112–24.

\item[MU] K.Murakami, T. Uno, Efficient algorithms for dualizing large-scale hypergraphs, Disc. Appl. Math. 170 (2014) 83-94.
 \item[RW] C.Rohwer, M.Wild, LULU Theory, idempotent stack filters, and the mathematics of vision of Marr, Advances in Imaging and Electron Physics 146 (2007) 57-162.
    \item[W1]  M.Wild, Compression with wildcards: From CNFs to orthogonal DNFs by imposing the clauses one-by-one, The Computer Journal, Vol. 65 (2022) p.1073-1087.
     \item[W2]  M.Wild, The joy of implications, aka pure Horn formulas: Mainly a survey, Theoretical Computer Science 658 (2017) 264-292.
     \item[ W3]  M.Wild, Compression with wildcards: Abstract simplicial complexes, Quaestiones Mathematicae 46 (2023) 1151-1173.
      \item[ W4]  M.Wild, Compression with wildcards: All exact or all minimal hitting sets,
    Open Mathematics 2023; 21:20220596. 
     
\end{itemize}

\end{document}